\documentclass[twocolumn,graphics,floatfix,a4paper,prb]{revtex4}
\usepackage{graphicx, epsfig}
\usepackage{amsmath, amssymb}
\usepackage{color}
\usepackage{braket}
\usepackage{fancyhdr}
\usepackage{bbm}

\newcommand{\Fig}[1]{Fig.~\ref{#1}}

\newcommand{\comment}[1]{\textit{}}
\newcommand{\bit}{\begin{itemize} \setlength{\itemsep}{0ex} \setlength{\topsep}{0ex} } 
\newcommand{\eit}{\end{itemize}}
\newcommand{\be}{\begin{equation}}
\newcommand{\ee}{\end{equation}}
\newcommand{\bea}{\begin{eqnarray}}
\newcommand{\eea}{\end{eqnarray}}
\newcommand{\ba}{\begin{align}}
\newcommand{\ea}{\end{align}}
\newcommand{\SKIP}[1]{}

\def \w{\omega}
\def \ra{\rightarrow}
\def \ua{\uparrow}
\def \da{\downarrow}

\def \e{\varepsilon}
\def \tw{\epsilon_{tw}}

\begin{document}
\title{Efficient DMFT impurity solver using real-time dynamics with Matrix Product States}
\author{Martin Ganahl}
\affiliation{Institute of Theoretical and Computational Physics, Graz University of Technology, 8010 Graz, Austria}
\affiliation{Perimeter Institute for Theoretical Physics, 31 Caroline Street North, Waterloo, ON N2L 2Y5, Canada}
\author{Markus Aichhorn}
\affiliation{Institute of Theoretical and Computational Physics, Graz University of Technology, 8010 Graz, Austria}
\author{Patrik Thunstr\"om}
\affiliation{Institute of Solid State Physics, Vienna University of Technology, 1040 Vienna, Austria}
\author{Frank Verstraete}
\affiliation{Faculty of Physics, University of Vienna, Boltzmanngasse 5,  1090 Vienna, Austria}
\affiliation{Department of Physics and Astronomy, Ghent University, Ghent, Belgium}
\author{Karsten Held}
\affiliation{Institute of Solid State Physics, Vienna University of Technology, 1040 Vienna, Austria}
\author{Hans Gerd Evertz}
\affiliation{Institute of Theoretical and Computational Physics, Graz University of Technology, 8010 Graz, Austria}

\begin{abstract}
We propose to calculate spectral functions of quantum impurity models using the Time Evolving Block Decimation (TEBD) for Matrix Product States.
The resolution of the spectral function is improved by a so-called linear prediction approach.
We apply the method as an impurity solver within the Dynamical Mean Field Theory (DMFT) for the 
single- and two-band Hubbard model on the Bethe lattice. For the single-band model we observe sharp features
at the inner edges of the Hubbard bands. A finite size scaling shows 
that they remain present in the thermodynamic limit. 
We analyze the real time-dependence of the double occupation after adding a single electron and observe oscillations at the 
same energy as the sharp feature in the Hubbard band,
indicating a long-lived coherent superposition of states that correspond to the Kondo peak and the side peaks.
For a two-band Hubbard model we observe
an even richer structure in the Hubbard bands, which cannot be related
to a multiplet structure of the impurity, in addition to 
sharp excitations at the band edges of a type similar to the single-band case.
\end{abstract}

\maketitle

\section{introduction}
The field of strongly correlated materials has experienced vast growth during the last 
three decades. 
Electronic correlations pose a particular challenge for theory, and many phenomena such as high-temperature superconductivity have eluded a proper understanding to this day. 
A correlation phenomenon that is non trivial but nonetheless fully understood is the Kondo effect \cite{kondo_resistance_1964,wilson_renormalization_1975}. 
The Kondo model and its cousin, the Anderson impurity model, are not only relevant for its original purpose, i.e., for  magnetic impurities in solids, 
but also for quantum dots \cite{Kouwenhoven01a} and
even for bulk materials. For the latter, dynamical mean-field theory (DMFT) 
\cite{metzner_correlated_1989,georges_dynamical_1996,georges_strongly_2004}
maps a bulk lattice model onto the self-consistent solution of an Anderson impurity model
and includes in this way a major part of electronic correlations, namely the local ones.

Strong electronic correlations make the computational 
cost of directly solving the Schr\"odinger equation prohibitively large, and advanced numerical methods, 
often approximate ones, are in many cases the only viable option. At the forefront of these methods
lie the Quantum Monte-Carlo (QMC) technique \cite{assaad_world-line_2008,gull_continuous-time_2011}, the Numerical Renormalization Group 
\cite{bulla_numerical_2008,wilson_renormalization_1975}, the Density Matrix Renormalization Group (DMRG) 
\cite{white_density_1992,white_density-matrix_1993,schollwock_density-matrix_2011}, cluster approaches like Cluster Perturbation Theory \cite{senechal_spectral_2000} 
and the Variational Cluster Approach \cite{potthoff_variational_2003}
and DMFT \cite{metzner_correlated_1989,georges_dynamical_1996,georges_strongly_2004}.
All these methods have their strengths and weaknesses. QMC gives formally the exact solution, but is in practice 
plagued by statistical errors and the sign-problem. The NRG excels at capturing the low-energy physics, but has a hard time to resolve 
high-energy features in the spectrum and is restricted to impurity problems. The DMRG can treat both the low- and 
high-energy scale on equal footing, but is best suited for 1-dimensional (1d) models. Methods for higher 
dimensional (d$>$1) problems are scarce. One of the most promising among these is the DMFT \cite{georges_dynamical_1996} 
which becomes exact in the limit d$\ra \infty$
\cite{metzner_correlated_1989} and which yields an approximation for the 
finite dimensional lattice. The key quantity of DMFT is the 
{\it local lattice spectral function} 
$A(\w)$ which is calculated self-consistently. The framework of DMFT is readily established \cite{georges_dynamical_1996}, but the actual solution of the 
DMFT equations is complicated: it involves the calculation of the spectral function of an interacting impurity system which even for 
single-band models is highly non-trivial, and the complexity grows quickly with the number of considered bands (i.e. impurity orbitals).
Many different approaches have been proposed
to tackle the problem. The most common ones are QMC 
\cite{jarrell_hubbard_1992,rozenberg_mott-hubbard_1992,gull_continuous-time_2011,triqsnew,legendre,W2DYNAMCIS}, exact diagonalization (ED) 
\cite{caffarel_exact_1994,sangiovanni_static_2006,granath_distributional_2012,lu_efficient_2014}, and NRG \cite{zitko_energy_2009,pruschke_hunds_2005,bulla_zero_1999}.
QMC can efficiently handle multiple bands, but when formulated in imaginary time, it lacks high resolution of the spectral function. This is mainly
attributable to the ill-conditioned analytic continuation from imaginary to real frequencies. 
Based on the work in Refs.\ \cite{muhlbacher_real-time_2008,werner_diagrammatic_2009}, the QMC method has recently been extended to the calculation of real-frequency spectra
using bold line methods \cite{prokofev_bold_2007,gull_bold-line_2010} (see Refs.\ \cite{cohen_greens_2014-1,cohen_greens_2014} and references therein).
The analytic continuation is then traded for the introduction of a sign-problem.
ED naturally works with real energies, but it is severely limited by the number of possible sites. This again
reduces the spectral resolution considerably. Recently, two extensions to the ED method have been put forward \cite{granath_distributional_2012,lu_efficient_2014}
which increase the possible resolution of the spectral function \cite{granath_coherent_2014}.
The NRG on the other hand, being {\it designed} for impurity problems, achieves
very good spectral resolution at small energies. But due to a necessarily logarithmic discretization \cite{bulla_numerical_2008} of the bath 
density of states (DOS), high energy features of the bath are increasingly hard to resolve, which is also likely to affect the fixed point 
of the DMFT-iteration.
In addition, NRG has an intrinsic exponential growth of complexity with the number of bands of the
underlying lattice model. More than two bands \cite{greger_emergence_2013,pruschke_hunds_2005} have so far been unfeasible. 
The DMRG on the other hand offers several ways to be used as an impurity solver within the DMFT.
One possible way is to employ the dynamical DMRG (DDMRG) \cite{jeckelmann_dynamical_2002,kuhner_dynamical_1999} to obtain 
the DMFT-spectral function \cite{peters_spectral_2011,karski_single-particle_2008,karski_electron_2005,nishimoto_dynamical_2004},
whereas other approaches use a continued fraction expansion of the Greens function \cite{garcia_dynamical_2004,hallberg_density-matrix_1995}.
The broad application of DDMRG as an impurity solver is though hindered by the fact that one has
to perform a separate DMRG run for each frequency and to perform the inversion of an ill-conditioned 
system of linear equations, which can become very time consuming.
Recently, the Chebyshev expansion technique \cite{holzner_chebyshev_2011,weise_kernel_2006} 
has been proposed by some of us \cite{ganahl_chebyshev_2014} as an impurity solver for DMFT. 
Advantages are that it works at zero temperature, that no inversion problem has to be solved, 
and that the spectral function can be calculated with uniform resolution and high precision for all $\w$ in a single run. This reduces computational costs 
considerably, but on the other hand it is not straightforward to parallelize the method. In the present work
we propose to employ the Time Evolving Block Decimation (TEBD) \cite{vidal_efficient_2004,vidal_efficient_2003} for Matrix Product States 
(MPS) \cite{schollwock_density-matrix_2011,verstraete_matrix_2008} to compute the spectral function.
We combine it with a so called linear prediction technique \cite{white_spectral_2008,barthel_spectral_2009,ganahl_chebyshev_2014} to improve on the 
spectral resolution. Our method shares the advantages of the Chebyshev approach, but it can be carried out without the need of
explicitly adding states. By employing a Suzuki-Trotter decomposition it can easily be parallelized and is therefore both a very 
precise and very efficient method. We note that the numerical costs of the Chebyshev technique and time evolution both scale as $(d\chi)^3$,
where $d$ is the local Hilbert space dimension and $\chi$ is the matrix dimension employed.
The prefactor however
depends strongly on the parameters used in the two approaches. For a recent improvement of the Chebyshev technique see \cite{wolf_spectral_2015}. For the case
of the single-band Hubbard model, the run time of the time evolution method in our implementation was smaller by roughly an order of magnitude.

We apply it as a DMFT impurity solver for the single and the two-band Hubbard model on the $z\ra\infty$ Bethe lattice.

\section{Models}

\subsection{One-band Hubbard Model}
In this work we adress the computation of the local Greens function of the Hubbard model on the $z\ra\infty$ Bethe lattice using DMFT. In the limit of no interaction the model has a semi circular 
density of states (DOS) with a bandwidth of $2D$.
The central ingredient in DMFT 
is the iterative calculation of the spectral function $A_{\sigma,imp}(\w)=\bra{\Phi_0}c_{0\sigma}\delta(\w-H)c_{0\sigma}^{\dagger}+c_{0\sigma}^{\dagger}\delta(\w-H)c_{0\sigma}\ket{\Phi_0}$ 
of an impurity model. We consider the Single Impurity Anderson Model (SIAM)
\begin{equation}\label{eq:siam}
  H= \e_f\!\sum_{\sigma}\!n_{0\sigma}+U n_{0\downarrow}n_{0\uparrow}+\sum_{\kappa\neq 0,\sigma}\!\e_{\kappa} n_{\kappa\sigma}+\!\sum_{\kappa\neq 0,\sigma}\!V_{\kappa} c_{0\sigma}^{\dagger}c_{\kappa\sigma}^{\phantom{\dagger}}+h.c.\; ,
\end{equation}
where $c_{0\sigma},c_{0\sigma}^{\dagger}$ are fermionic annihilation and creation operators of spin $\sigma$ at the impurity site, 
$U$ and $\e_f$ are the interaction and the local potential 
of the underlying lattice model, i.e. the single-band Hubbard model, and $\e_{\kappa}$ and $V_{\kappa}$ are variational parameters which are optimized during the DMFT cycles
(see below). We assume a spin-symmetric coupling $V_{\kappa}$ and bath-dispersion $\e_{\kappa}$, and work at particle-hole symmetry $\e_f=-\frac{U}{2}$, in which case $A_{\sigma,imp}(\w)$ is
independent of spin. We thus omit the spin variable henceforth and simply write $A_{imp}(\w)$.

Greens and spectral functions can be calculated from the real time evolution and subsequent Fourier transformation of
\begin{align}\label{eq:greenfun}
  A_{imp}(t)=\frac{1}{2\pi}\bra{\Phi_0}\{c_0(t),c_0^{\dagger}(0)\}\ket{\Phi_0}
\end{align}
to
\begin{align}
  A_{imp}(\w)=\frac{1}{\sqrt{2\pi}}\int_{-\infty}^{\infty} dt e^{-i\w t}A_{imp}(t).
\end{align}
Here $c_0(t)$ is given in the Heisenberg picture, and $\ket{\Phi_0}$ is a non-degenerate ground state at zero energy.
Due to the hermiticity of $H$, the function $A_{imp}(\w)$ is real and normalized to unity.
One way to obtain $A_{imp}(\w)$ is to calculate the two quantities
\begin{align}\label{eq:GbGs}
  G^>(t)\equiv \bra{\Phi_0}c_0e^{-iHt}c_0^{\dagger}\ket{\Phi_0}\nonumber\\
  G^<(t)\equiv \bra{\Phi_0}c_0^{\dagger}e^{iHt}c_0\ket{\Phi_0}
\end{align}
for $-\infty<t<\infty$ and Fourier-transform them.
Using
\begin{align}\label{eq:Gbherm}
(G^<(t))^*=G^<(-t)\nonumber\\
G^>(t)=(G^<(t))^*,
\end{align}
where the second line is valid only at particle hole symmetry, the
spectral function can be expressed as 
\bea
A_{imp}(\w)=\frac{1}{(2\pi)^{3/2}}\int_{-\infty}^{\infty}dt \left(G^<(t)+G^<(-t)\right)e^{-i\w t}\nonumber\\
=\frac{2}{(2\pi)^{3/2}}\int_{-\infty}^{\infty}dt \Re(G^<(t))e^{-i\w t}.
\eea
To obtain the spectral function at particle-hole symmetry, we thus only need to
calculate $G^>(t)$ for $t>0$, as the real part of $G^>(t)$ is even in $t$.

\subsection{Two-band Hubbard Model}
The most promising feature of the method is its applicability to multi-band systems. We will demonstrate 
this by applying it as an impurity solver for the symmetric two-band Hubbard model \cite{pruschke_hunds_2005,koga_stability_2002,han_multiorbital_1998,held_microscopic_1998} 
on the Bethe lattice. Under the assumption that the bath is spin
symmetric the resulting effective impurity model assumes the form
$H=H_\textrm{loc} + H_\textrm{bath}$ with
\begin{widetext}
\begin{subequations}\label{eq:twoorbsiam}
\begin{align}
  H_\textrm{loc}&=\sum_{m\sigma} \e_m
  n_{m\sigma}+\frac{U}{2}\sum_{m\sigma}n_{m\sigma}n_{m\bar\sigma}+\frac{U'}{2}\sum_{m\sigma}n_{m\sigma}n_{\bar m\bar\sigma}
   +\frac{(U'-J)}{2}\sum_{m\sigma}n_{m\sigma}n_{\bar m\sigma}\nonumber\\
   &-J(c_{00\ua}^{\dagger}c_{00\da}c_{10\da}^{\dagger}c_{10\ua}+h.c.)-J(c_{00\ua}^{\dagger}c_{00\da}^{\dagger}c_{10\ua}c_{10\da}+h.c)\\
  H_\textrm{bath}&=\sum_{m\kappa\sigma}(V_{m\kappa}c_{m0\sigma}^{\dagger}c_{m\kappa\sigma}+h.c.)+\sum_{m\kappa\sigma}\e_{m\kappa}n_{m\kappa\sigma}
\end{align}
\end{subequations}
\end{widetext}
with $U'=U-2J$, and we choose $J=U/4$ throughout. Here, $c_{m0\sigma}$ is a fermionic annihilation operator of the correlated orbital $m\in\{0,1\}$ at 
the impurity site, $n_{m\sigma}$ are the corresponding particle number operators, and $c_{m\kappa\sigma}$ and $n_{m\kappa\sigma}$ are 
the bath-electron annihilation and particle number operators for $\kappa \geq 1$, respectively. Symbols $\bar\sigma$ and $\bar m$ denote 
complementary variables, e.g. if $\sigma=+1/2$, then $\bar\sigma=-1/2$, and similar for $m$.
Note that we use a fully rotational invariant interaction with all spin-flip and pair-hopping terms included.

\section{Methods}

\subsection{Dynamical mean field theory}
In this paper we address the single- and two-band Hubbard model on the $z\ra\infty$ Bethe lattice. This is convenient for two reasons:
(i) the DMFT yields exact results in this case, and (ii) the DMFT self-consistency scheme is especially simple.
The quantity of genuine interest in DMFT is the local lattice spectral function $A_{latt}(\w)$ of an interacting,
$d$-dimensional lattice problem (e.g the Hubbard model on the Bethe lattice).
At convergence it is identical to the impurity spectral function $A_{imp}(\w)$.
The basic idea of DMFT is to mimic the effect of the interacting lattice surrounding a given site by a suitably chosen bath
of free electrons. Interacting lattice site and surrounding bath yield an impurity problem described by Eq.(\ref{eq:siam}) or Eq.(\ref{eq:twoorbsiam}).
The bath can be represented by the hybridization function
$\Delta_m(\omega)=\sum_{\kappa}\frac{|V_{m\kappa}|^2}{\omega+ i\eta-\epsilon_{m\kappa}}$, with 
an imaginary part $\tilde \Delta_m(\w)\equiv-\frac{1}{\pi}\Im(\Delta_m(\w))=\sum_{\kappa\sigma}|V_{m\kappa}|^2\delta(\w-\e_{m\kappa})$.
The general outline of the DMFT cycle is as follows: For each correlated orbital $m$ on the impurity site we initially guess a 
$\tilde \Delta_m^{n=0}(\w)$, where $n$ is an iteration index. 
A set of SIAM parameters $V_{m\kappa}$ and $\e_{m\kappa}$ is then obtained by discretizing $\tilde \Delta_m^{n=0}(\w)$
as described in Ref.\cite{bulla_numerical_2008}. We use a discretization scheme linear in energy in this work.
The method can deal with any discretization. After the discretization the system has a linear chain geometry 
\begin{equation}\label{eq:siamdisc}
  H= H_\textrm{loc} 
  +\sum_{mi\sigma}(t_{mi\sigma}c_{mi\sigma}^{\dagger}c_{mi+1\sigma}^{\phantom{\dagger}}+h.c.)+\sum_{mi\sigma}\e_{mi\sigma}n_{mi\sigma}
\end{equation}
with only nearest neighbor hopping $t_{mi}$ and local potentials $\e_i$. The number of discretization parameters corresponds to the chain length $N$. 
Using an impurity solver we calculate the impurity spectral function $A^n_{imp,m}(\w)$ of Eq.(\ref{eq:siamdisc}), from which one
obtains the new 
\be
\tilde \Delta_m^{n+1}(\w)=\frac{D^2}{4}(\alpha A^n_{imp,m}(\w) + (1-\alpha)A^{n-1}_{imp,m}(\w))
\ee
with a mixing parameter $\alpha\in[0,1]$ (``underrelaxation'') that can be adjusted for faster convergence. Here, $2D$ is the bandwidth of the 
non-interacting spectral function of the Bethe lattice. In the following, all results are 
given in units of $D$.
$\tilde\Delta(\w)_m^{n+1}$ is then again discretized, and the loop is iterated upon convergence,
i.e. until
\be
A^{n+1}_{imp,m}(\w)=A^n_{imp,m}(\w)=A_m^{latt}(\w)\equiv A_m(\w).
\ee
For our calculations we enforce particle-hole and spin symmetry which results in $\e_{i\sigma}=0$ in Eq.(\ref{eq:siamdisc}).

For the two-band Hubbard model we focus on the symmetric model in which at $U/D=0$ both bands have the same bandwidth $2D$. 
For the discretization we use a symmetric setup with two DMFT-baths of lengths $N_1=N_2$. Each bath contains electrons of up and 
down spin character.

The two-band model has been frequently investigated in the past \cite{greger_emergence_2013,pruschke_hunds_2005,held_microscopic_1998}.
Like the single-band model it exhibits a MI transition at a finite $U_c/D\approx 2.2$, for $J=U/4$.\cite{pruschke_hunds_2005}
 
Using the NRG as an impurity solver \cite{pruschke_hunds_2005}, 
it has been observed that a finite $J$ leads to a strong renormalization of the Kondo temperature and hence affects the MI transition. 
Obtaining accurate results, especially for the high energy features of the Hubbard bands remains a challenging problem.

\subsection{Time Evolving Block Decimation}
The essential task of the impurity solver in DMFT is to calculate the greater (or lesser) Greens function $G^>(t)=\bra{\Phi_0}c_0e^{iHt}c_0^{\dagger}\ket{\Phi_0}$. We achieve this
by calculating the ground state $\ket{\Phi_0}$ using the DMRG and subsequently employ the Time Evolving Block Decimation (TEBD) \cite{vidal_efficient_2003,vidal_efficient_2004}
to evolve $c_0^{\dagger}\ket{\Phi_0}$ forward in time. We use a second order Trotter breakup with $\Delta t D=0.00625$, and in the single-band case measure $G^>(t_nD)$ every 25 Trotter steps,
thus $t_n D=0.15625\;n$, with $ n\in\{0,\dots,N_{t,max}-1\}$.
For the two-band model we split the calculation of the Greens function into two runs, one forward and one backward in time,
which allows us to reach longer times and/or larger system sizes with a smaller computational effort \cite{barthel_unknown_2013}. 
From these runs we calculate the Greens function
on a grid with $N_{t,max}$ grid-points $t_n D=0.375\;n$, with $\Delta t D=0.00625$. We then apply 
the linear prediction method (see below) to extrapolate 10000 further points. 
In the single-band case we also employ an unfolding procedure \cite{saberi_matrix-product-state_2008} to separate the spin degrees of freedom of the electrons into a left chain 
containing up spins and a right chain containing down spins which is computationally more efficient. In this case, the total length of the system is $2N$, that is each bath
has a length of $N-1$ sites.
A decoupling for two and more bands can be carried out in a similar spirit with multiple chains connected at the impurity site in a star-like geometry \cite{holzner_matrix_product_2010}
where every chain carries a band and spin index. For the two-band case however, we only applied a decoupling of the orbital degrees of freedom. The impurity orbital 
is in this case located at the center of a chain, where the left $N_1$ sites contain the $m=0$ bath and impurity orbital, and the right $N_2$ sites the $m=1$ bath and impurity orbital.
Each of the $N_1-1$ and $N_2-1$ bath orbitals contains electrons of up- and down-spin flavour. 

In the TEBD, only a limited number $\chi$ of Schmidt-states $\ket{\lambda_\alpha}$ and Schmidt-values $\lambda_{\alpha}$ can be kept at a certain bond connecting two sites, which is the
major approximation of the method. The error of a single time step can be quantified by the truncated weight \cite{vidal_efficient_2003,schollwock_density-matrix_2011}
\be
\tw=1-\sum_{\alpha=\chi+1}^{d\chi}\lambda^2_{\alpha}
\ee
obtained after truncating the state down to a matrix dimension $\chi$, where $d$ is the local Hilbert space dimension.
In our implementation of the TEBD, after every time step the matrix dimension is reduced just enough to obtain the prespecified $\tw$.
Additionally, we set a hard limit for the maximum matrix dimension of $\chi=500$ or $750$ in the single-band and $\chi=800$ or $1000$ in two-band model.

\subsection{Linear Prediction}
The so called linear prediction technique \cite{numrep_2007,ganahl_chebyshev_2014,white_spectral_2008,barthel_spectral_2009} is a very simple and powerful method for the extrapolation of time series.
It amounts to describing the time series as a sum of many exponentials or, equivalently, the spectrum as a sum of many Lorentzians.
On the basis of $N_{t,max}\equiv 2N_t$ calculated data points $\{x_i\}$, $1\le i \le 2N_t$ at equidistant times $t_i$, one predicts data points
for $t_n,n>2N_t$ as a linear combination of the first $N_t$ data points:
\begin{align}
x_n \approx \tilde{x}_n\equiv-\sum_{j=1}^{N_t} a_j x_{n-j}.\label{eq:prediction}
\end{align}
One obtains the optimal $\{a_j\}$ by minimization of a cost function
\begin{align}
  \mathcal{F}=\sum_{n=N_t+1}^{2N_t} |\tilde x_n-x_n|^2 \label{eq:linpred},
\end{align}
which yields
\begin{align}\label{eq:linpred_matrix}
 R{\bf a}&=-{\bf r},\\ \nonumber
  R_{ij}&=\sum_{n=N_t+1}^{2N_t}  w_n x^*_{n-i}x_{n-j},&\quad r_i=\sum_{n=N_t+1}^{2N_t} w_n x^*_{n-i}x_n,
\end{align}
for $1\le i,j \le N_t$. Eq. (\ref{eq:linpred_matrix}) is inverted using a pseudo-inverse with cutoff $\delta$. 
Data points at $N_t+k$ ($k>0$) can then be predicted from
\begin{equation}
\tilde x_{N_t+k}= \sum_{n=1}^{N_t} [M^k]_{1\,n} \, x_{N_t+1-n} \,, \label{eq:predictlp}
\end{equation}
with
\[M=\left(\begin{array}{ccccc}
-a_1 & -a_2 & -a_3 & \dots & -a_{N_t}\\
1   & 0     &   0  & \dots & 0\\
0   & 1     &   0  & \dots & 0\\
\vdots   & \ddots  & \ddots &\ddots & \vdots\\
0  & 0       & \dots &1 & 0\\
\end{array}\right).
\]
All spectral functions have been obtained by predicting 10000 further data points on top of $N_{t,max}$ computed data points.
Due to the exponential dependence of $M$ in Eq.(\ref{eq:predictlp}), any eigenvalue $\lambda$ of $M$ that is larger
than unity has to be either renormalized to unity or set to zero, in order to avoid divergence in the prediction \cite{barthel_spectral_2009,ganahl_chebyshev_2014}.
The interplay of $\delta$ and the eigenvalue rescaling is investigated in more detail in the appendix. There
we show that zeroed eigenvalues yield better results.

The linear prediction algorithm has two parameters $N_{t,max}$ (``time window'') and $\delta$ (pseudo-inverse cutoff). From analysis of
the dependence of the DMFT-fixed point on these parameters (see appendix), we found $N_{t,max}=350$, and $\delta=10^{-6}$ or smaller, in
conjunction with setting large eigenvalues of the prediction matrix to 0, to yield good results. Unless stated otherwise,
these parameter values were used to obtain the results in this paper.

\section{Results}
To verify our approach we benchmarked our results for the impurity spectral function of a SIAM 
with results of the dynamical DMRG \cite{raas_spectral_2005}, to our knowledge the most precise data available (see appendix),
and found excellent agreement. In the following we present our results for the one and two-band Hubbard model.
 
\subsection{One-band Hubbard model} 
\begin{figure*}
  \includegraphics[width=2\columnwidth]{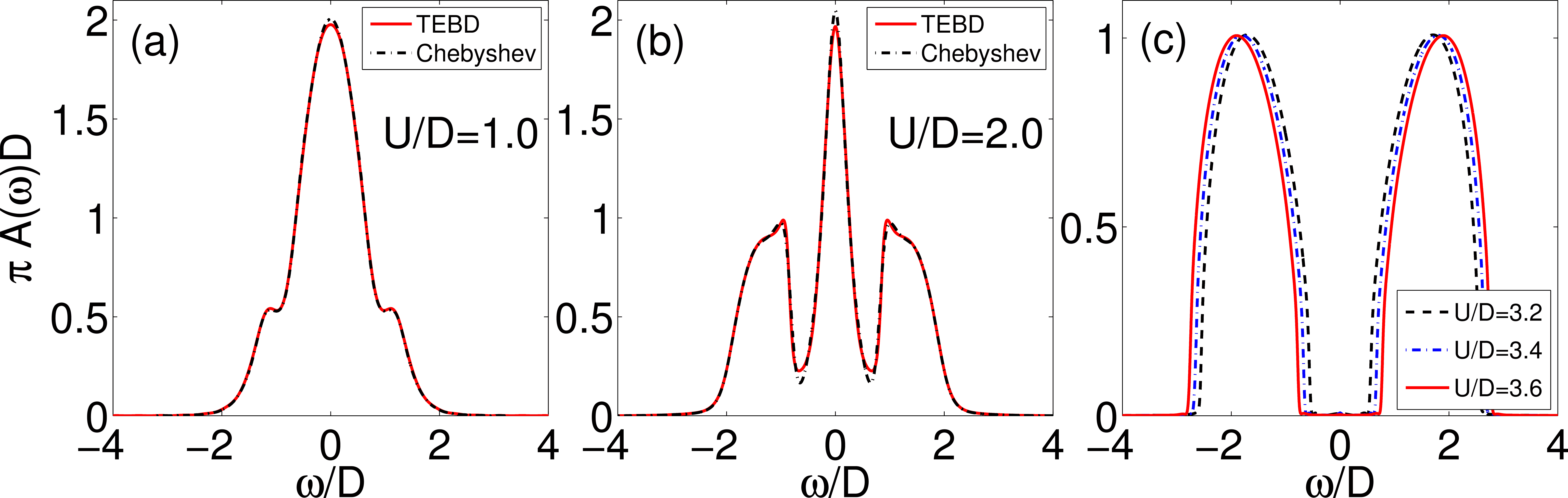}
\caption{(a) and (b): DMFT spectral function of the half-filled Hubbard model on the Bethe lattice for $U/D=1.0,2.0$ ($N=120,N_{t,max}=200,\chi=500,\tw=10^{-10}$) as
  obtained from TEBD (red solid line). For comparison we plot results obtained using the Chebyshev expansion method \cite{ganahl_chebyshev_2014} (black dash-dotted line).
  (c) DMFT spectral functions in the insulating phase for $U/D=3.2,3.4,3.6$. For better convergence we used the modified update scheme with $\alpha=0.3$ in (c).
  Other parameters as in (a) and (b).}\label{fig:smallu_large_u}
\end{figure*}

We start by applying our method to DMFT for the single-band Hubbard model on the Bethe lattice for interaction strengths
$U/D=1.0$ and $U/D=2.0$ in the metallic region and $U/D=3.2,3.4$ and $3.6$ in the insulating region.
We use $N_{t,max}=200$. \Fig{fig:smallu_large_u}(a) and (b) show the results for $U/D=1.0$ and $2.0$ (red solid lines). At $U/D=2.0$, distinct features at the inner edges of the Hubbard 
satellites start to emerge, as has been observed in previous DDMRG 
\cite{karski_single-particle_2008,karski_electron_2005} and NRG \cite{zitko_energy_2009} studies, in QMC calculations \cite{gull_bold-line_2010} as well as in MPS calculations 
with Chebyshev moments \cite{ganahl_chebyshev_2014}, and recently also in advanced ED calculations \cite{granath_coherent_2014,lu_efficient_2014,granath_distributional_2012}.
For comparison we show results obtained with the Chebyshev expansion technique \cite{ganahl_chebyshev_2014} (black dash-dotted lines). They are
compatible with our present results. For values of $U/D=3.2,3.4$ and 3.6 we plot results in \Fig{fig:smallu_large_u}(c). In contrast to \Fig{fig:smallu_large_u}(a) 
and (b), we use a mixing parameter of $\alpha=0.3$ in \Fig{fig:smallu_large_u}(c), which yields a smoother convergence to the insulating solution. 
If no mixing is applied, the spectra alternate between an insulating and a metallic solution with a tiny quasi-particle weight.
This effect is particularly strong at $U/D=3.2$ close to the transition and is also enhanced when increasing $N_{t,max}$ 
or decreasing $\delta$, which we attribute to Trotter and truncation effects in the time series. We note that in 
\Fig{fig:smallu_large_u}(c) there is some small residual spectral weight left in the gap region (of the order of $10^{-3}$).

\subsection{Sharp peaks in the Hubbard bands}
We proceed to study the metallic state in the coexistence region
$U_{c1}/D \leq U/D \leq U_{c2}/D$, where $U_{c1}/D\approx 2.38$, and $U_{c2}/D\approx 3.0$ \cite{karski_single-particle_2008}.
For such $U/D$, the narrowing of the quasi-particle peak at $\w/D=0$ and the appearance of sharp side peaks in the Hubbard band 
\cite{ganahl_chebyshev_2014,lu_efficient_2014,granath_distributional_2012,karski_electron_2005,karski_single-particle_2008} 
make a high-resolution calculation of $A(\w)$ a challenging task.

\begin{figure}
  \includegraphics[width=1.0\columnwidth]{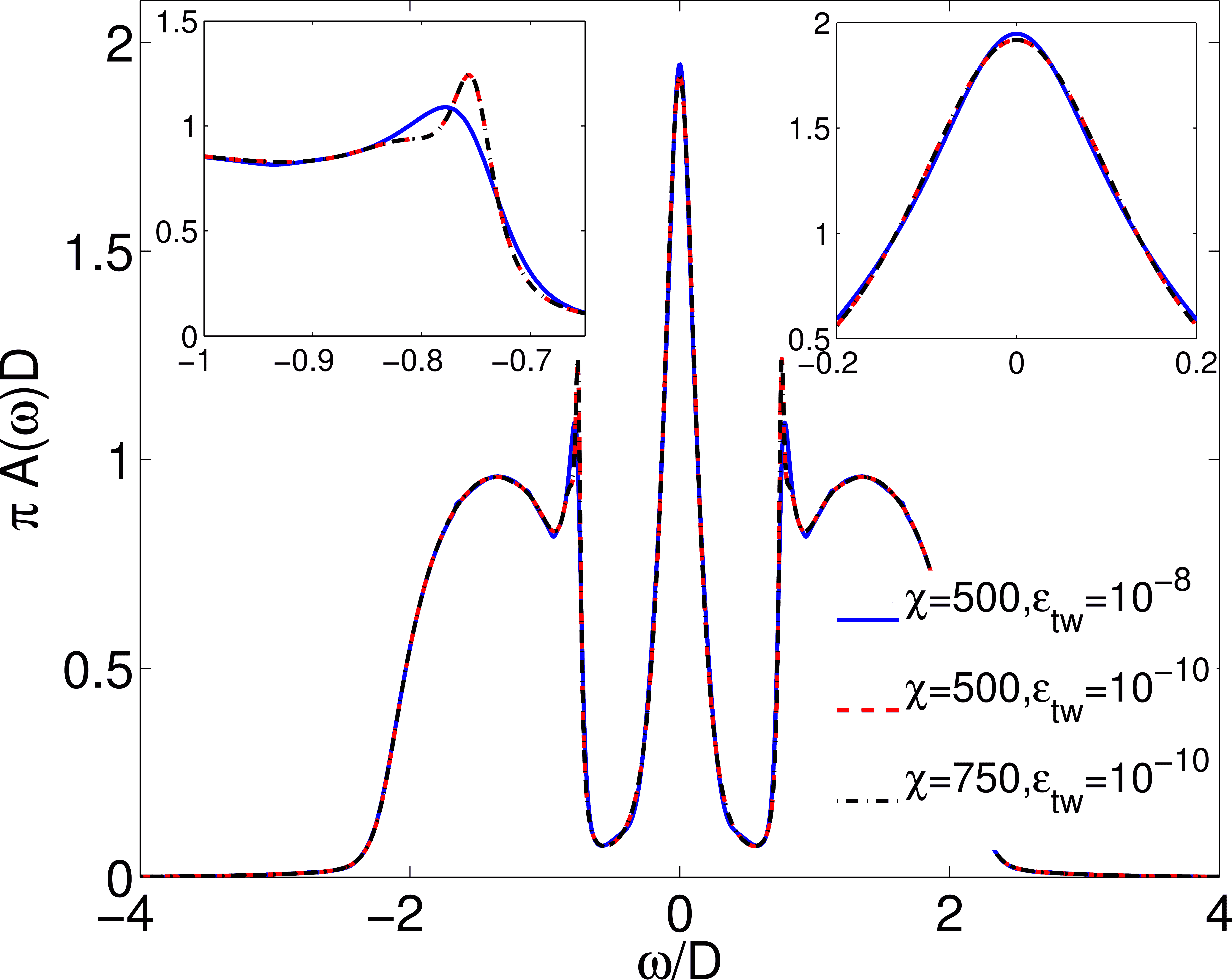}
\caption{DMFT spectral function for $U/D=2.4$, comparing three different computational parameter sets $\chi$ and $\tw$ ($N=150$). The spectral function shows the quasi-particle peak at
$\w/D=0$ and two broad Hubbard bands at $\w/D\approx\pm U/2D$. Additionally, at the inner edges of the Hubbard band we observe a sharp feature. Insets:
 closeups of the sharp peak at the inner side of the Hubbard bands (left) and the quasi-particle peak (right).}\label{fig:u2.4_trweight}
\end{figure}

In \Fig{fig:u2.4_trweight}, we present $A(\w)$ for $U/D=2.4$ and a chain length of
$N=150$ sites. We clearly observe a separation of energy-scales into a sharp resonance at $\w/D=0$ and two broad Hubbard satellites
at $\w/D\approx \pm U/2D$, decorated with two sharp features at the inner edges of the Hubbard peaks. The insets are closeups on the 
sharp side peak of the left Hubbard satellite (left) and the quasi-particle peak (right). 

It can be shown analytically~\cite{hewson_93} 
that the exact value of $\pi D A(\omega=0)=2.0$ for all values of $U/D<U_c/D$.
This pinning of the height of the Kondo-resonance is fulfilled to good accuracy in our simulations.
In \Fig{fig:u2.4_trweight} we analyze the convergence of the results with increasing precision of the MPS calculations. The latter is governed by the
truncation error $\epsilon_{tw}$, related to the MPS 
matrix-dimension $\chi$, which bounds the number of Schmidt states kept at each bi-partition during the simulation.
\Fig{fig:u2.4_trweight} shows results for three different simulations. With increasing precision we initially observe a sharpening of the Hubbard side-peaks. 
From the plot we conclude that using $\chi=500,\tw=10^{-10}$ already yields converged results. 
The magnetiztion of our DMRG ground state is zero to within $10^{-7}$ accuracy, which rules out that these sharp excitations are 
artefacts from any spurious magnetic ordering due to the DMRG truncation.
A slight decrease of the quality of the pinning criterion 
is most likely related to linear prediction inaccuracies. 

\begin{figure}
  \includegraphics[width=1.0\columnwidth]{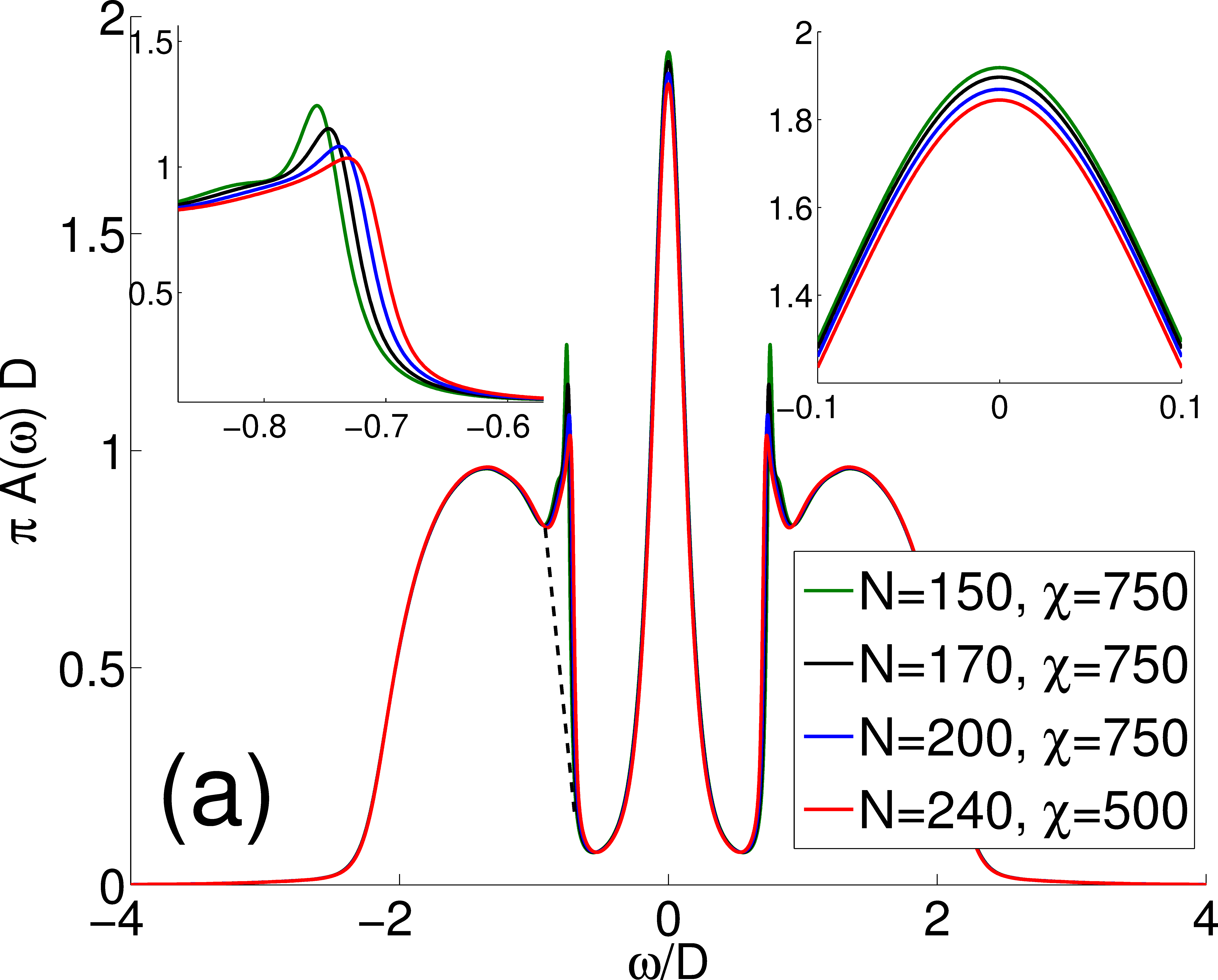}
  \includegraphics[width=0.49\columnwidth]{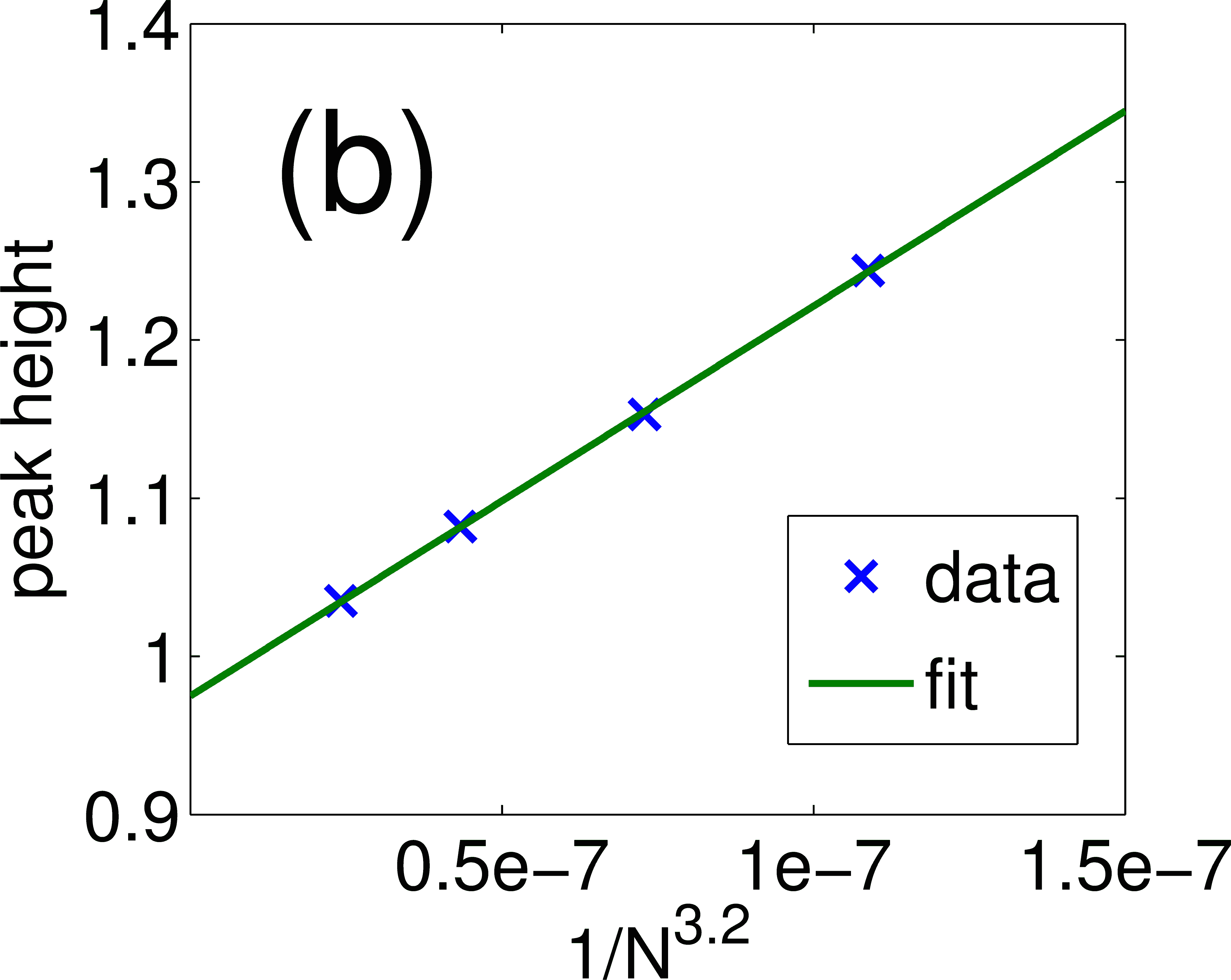}
  \includegraphics[width=0.46\columnwidth]{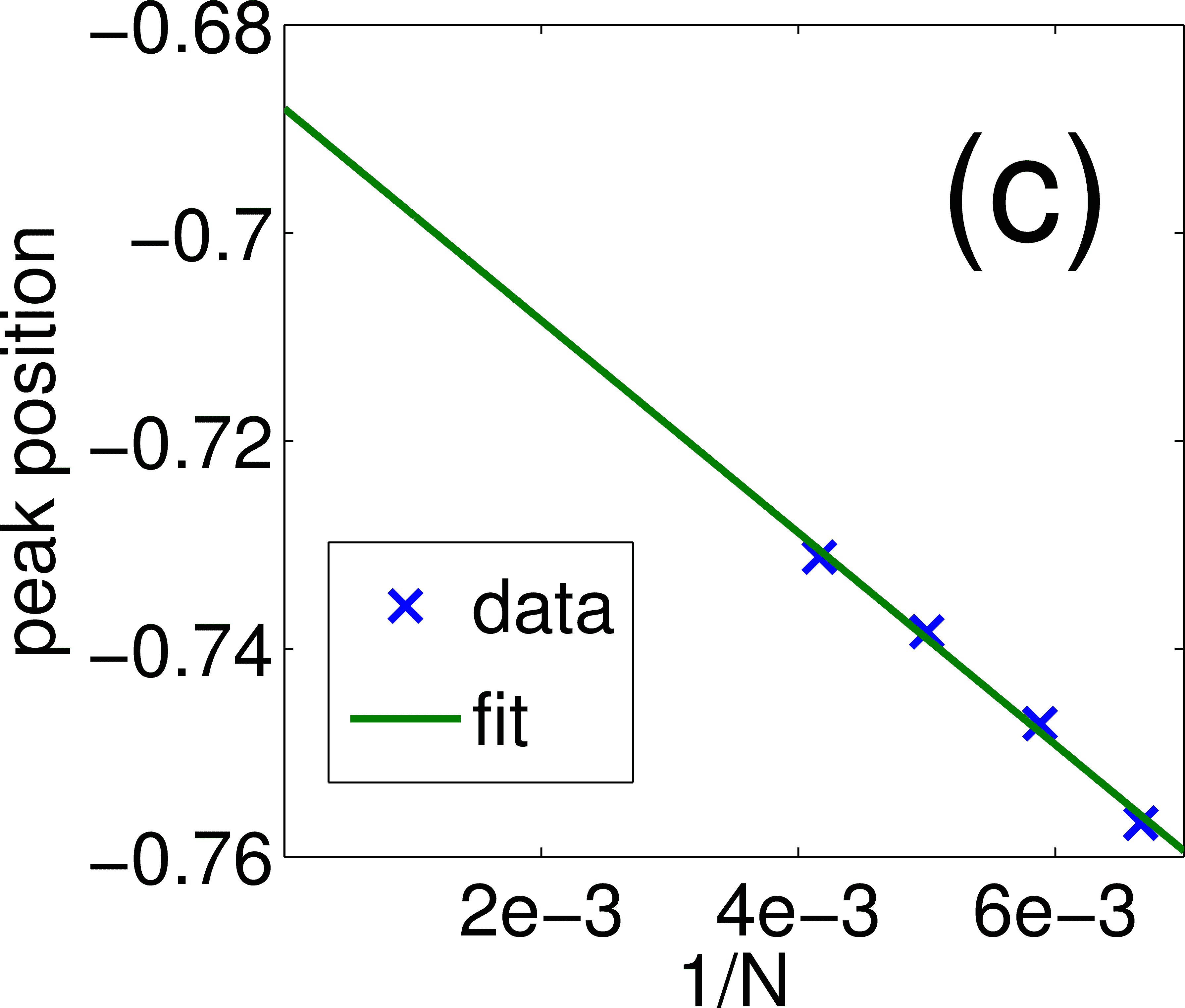}
  \includegraphics[width=0.49\columnwidth]{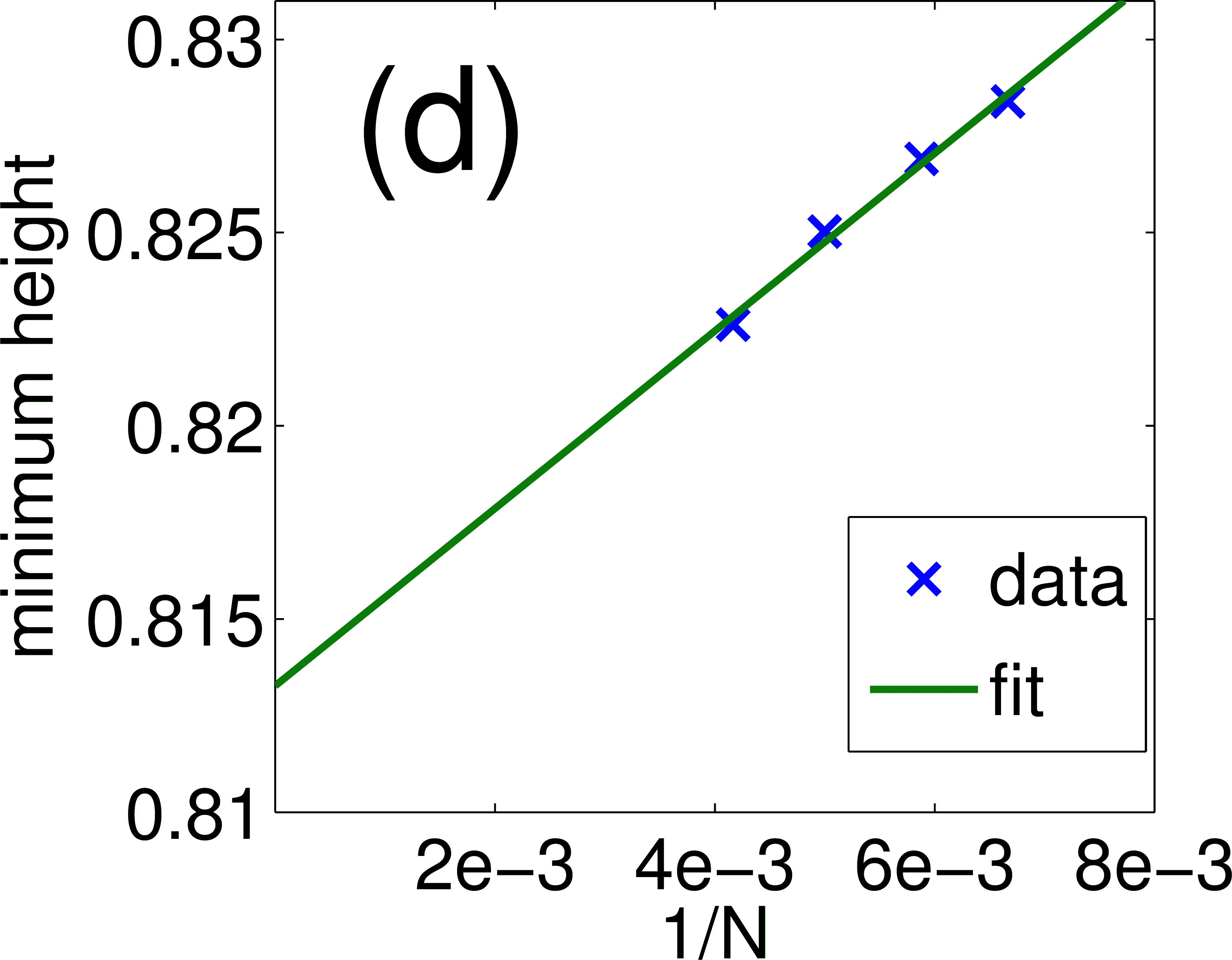}
  \includegraphics[width=0.49\columnwidth]{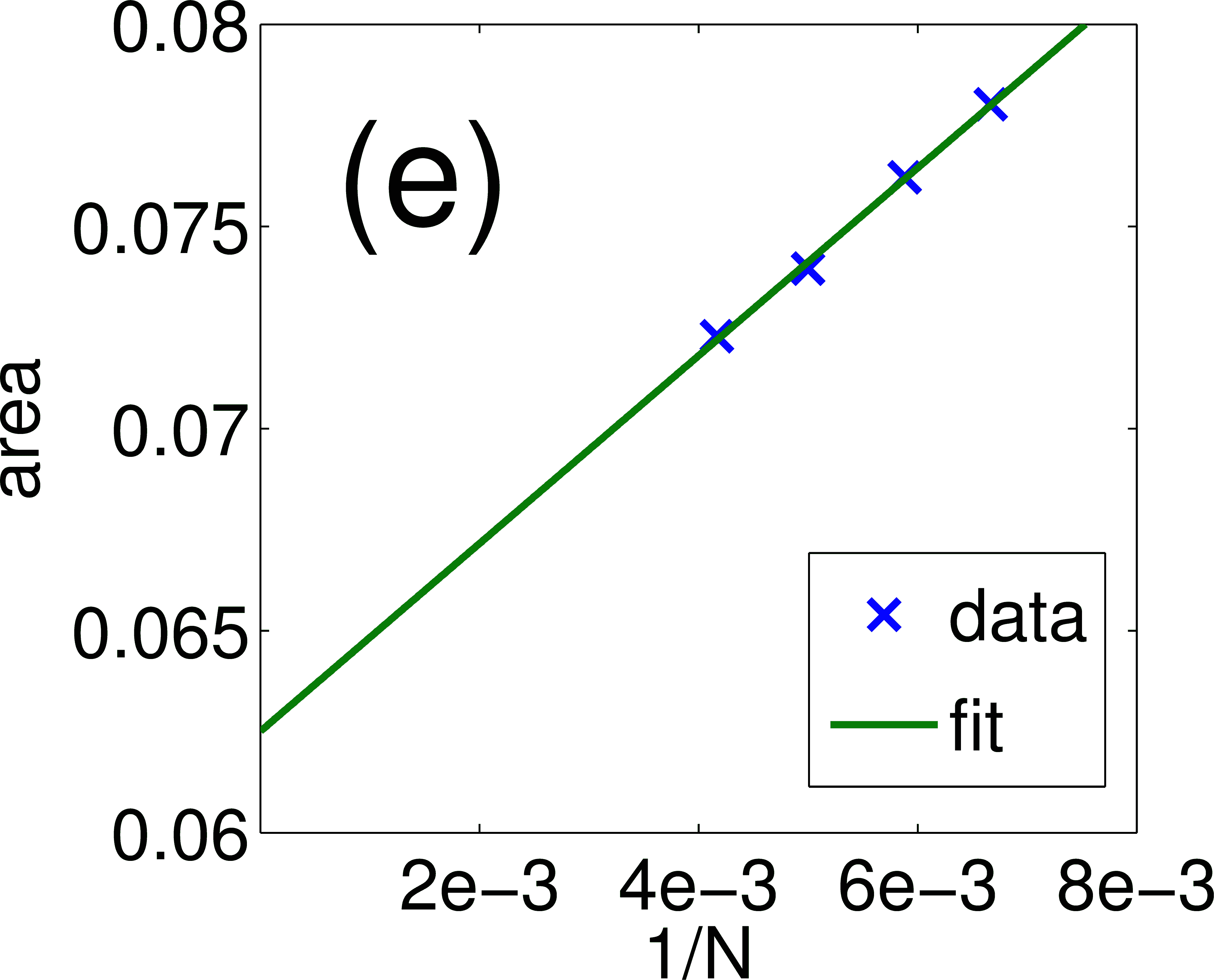}
\caption{(a) Spectral function of converged DMFT cycles at $U/D=2.4$ for different chain lengths $N=150,170,200,240$, ($\chi=500,750,\tw=10^{-10}$). Insets are 
closeups of Hubbard bands (left) and quasi-particle peak (right). Below: Finite size scaling with respect to $N$ of (b) the height of the inner side peak,
  (c) the position of the inner side peak, (d) minimum of the inner side peak and (e) area between the dashed black line and inner side peak in (a).
}\label{fig:u2.4_diffN}
\end{figure}

A second important parameter is the chain length $N$, which is directly related to the number of discretization points
of the bath spectral function. In \Fig{fig:u2.4_diffN}(a) we present DMFT-spectra for $U/D=2.4$ and different system sizes $N=150\dots 240$, with $\chi=500$ or $750$ and $\tw=10^{-10}$.
With increasing system size $N$, we observe a shift of the Hubbard side-peak position towards smaller $|\w/D|$ (left inset \Fig{fig:u2.4_diffN}), as well as a reduction of its height 
\cite{karski_single-particle_2008,karski_electron_2005}.
A similar reduction is observed in the height of the quasi-particle peak (right inset in \Fig{fig:u2.4_diffN}),
in violation of the pinning criterion. However, the DMFT
self-consistency cycle is stable and still converges, and the violation does not
increase during this cycle.
We note that the spectral height at $\w/D=0$ is the quantity which is most susceptible to small errors in the large time evolution in our method, 
more so than the spectrum at other frequencies.

In \Fig{fig:u2.4_diffN}(b) to (e) we display a finite size scaling of the side peaks, which demonstrates that they remain present in the infinite size limit.
\Fig{fig:u2.4_diffN}(b) shows a finite size scaling of the height of the side peak (the exponent $3.2$ was found to map the data to a straight line), and (c) 
a scaling of its position. In panel (d) we present a finite size scaling of the minimum to the left of the left side peak. 
Finally panel (e) shows a scaling of an approximate measure of the area of the side peak, namely the area between the left side peak and a straight 
line through the local minimum, tangent to spectrum at the global minimum (see dashed
black line in the lower Hubbard band of the main panel in \Fig{fig:u2.4_diffN}(a)).
Since the height of the central peak is also size-dependent, the exact properties of the side peaks are likely to change 
with higher precision of the calculations. The important point 
(further strengthened by a slight curvature of the data)
is that they converge to finite values, i.e. the side peaks remain present in the thermodynamic limit.

\begin{figure}
  \includegraphics[width=1.0\columnwidth]{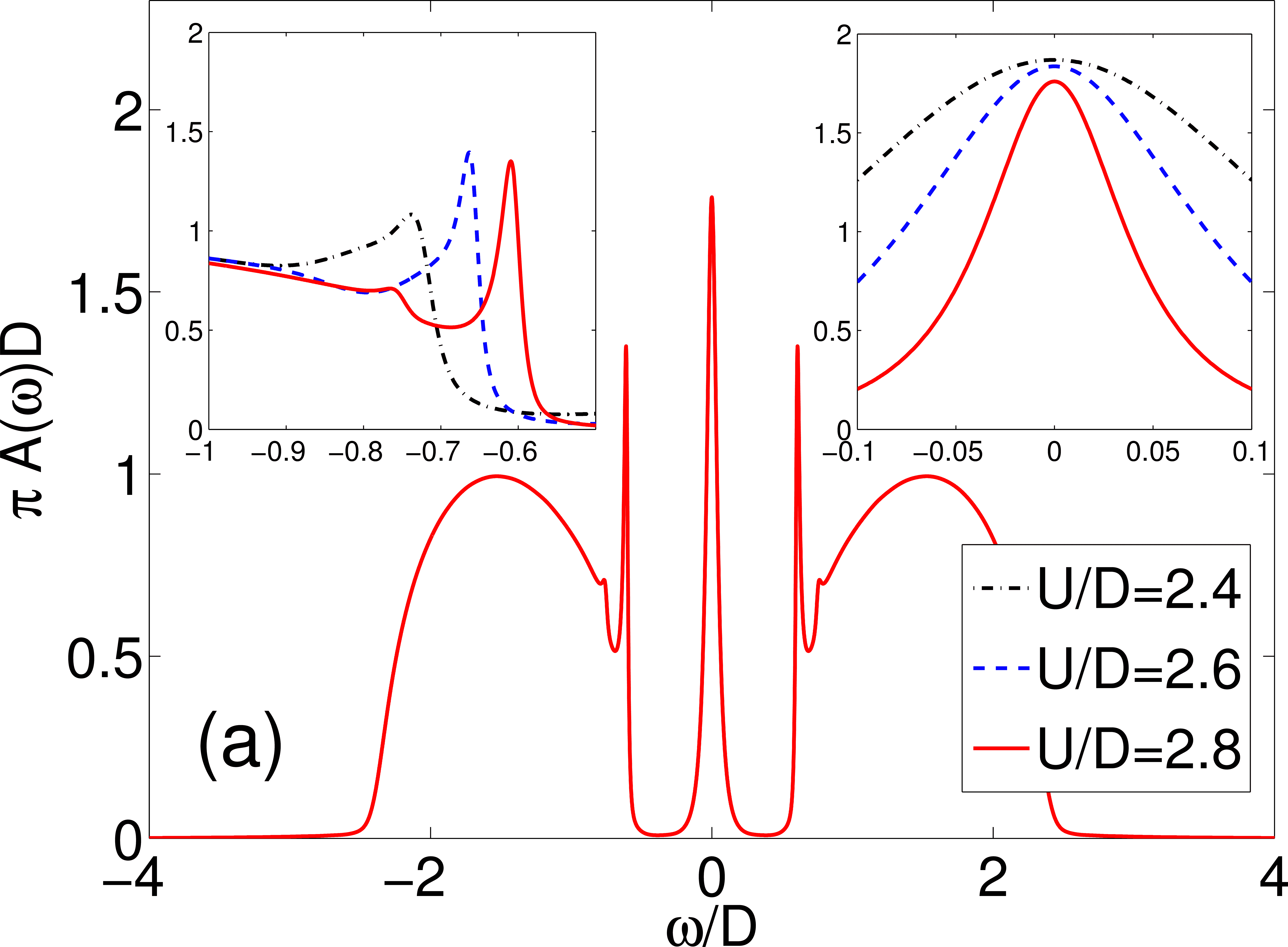}
  \includegraphics[width=1.0\columnwidth]{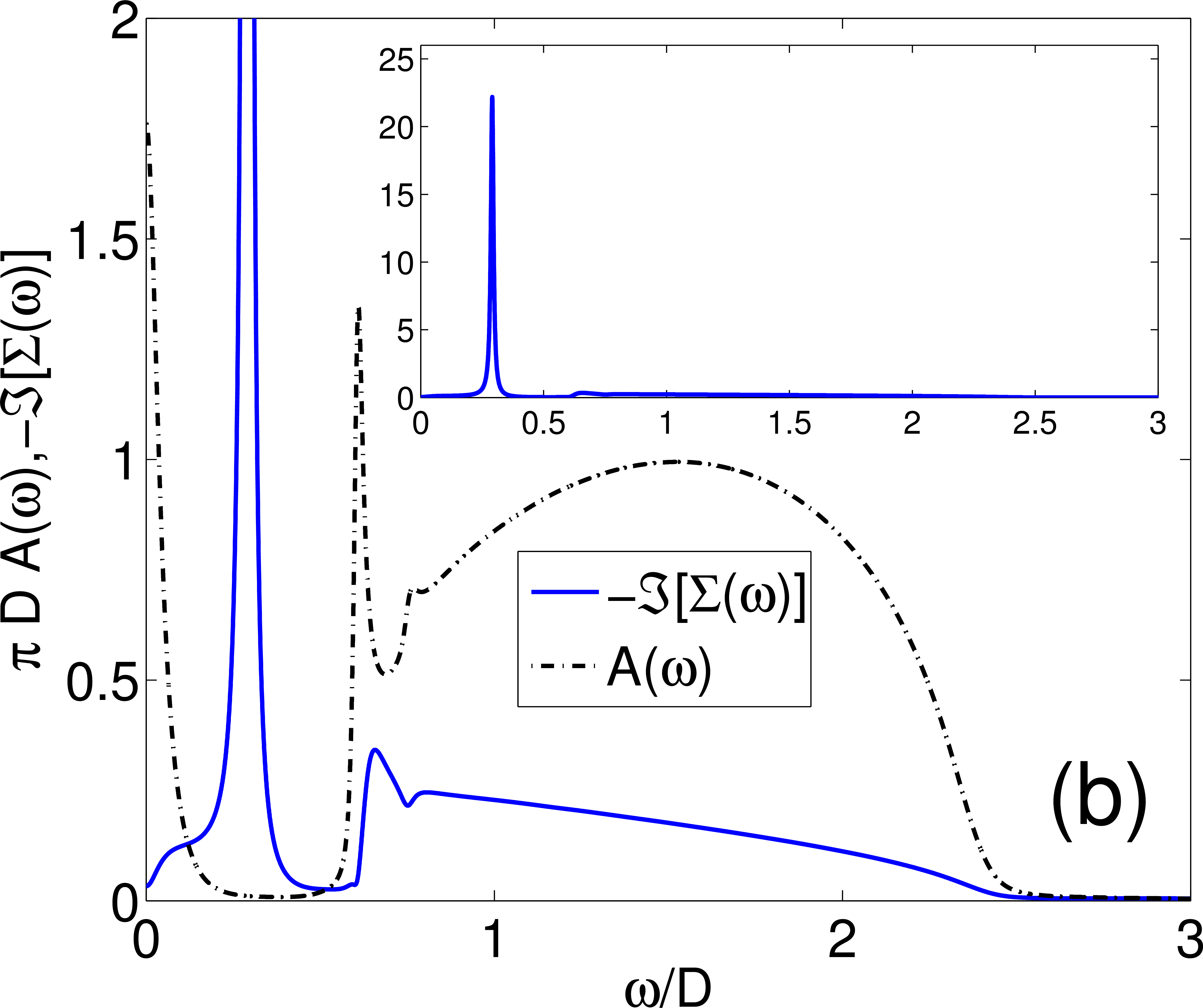}
\caption{(a) Spectral function for $U/D=2.8 ,(N=200, \chi=750,\tw=10^{-10})$. Right inset: Increasing $U/D=2.4,2.6,2.8$
we observe a narrowing of the quasi-particle peak. 
The sharp peaks at the inner side of the Hubbard bands get more pronounced
and are shifted towards smaller $|\w/D|$ (left inset). In the region between the quasi-particle peak and the Hubbard bands, the spectral weight is largely suppressed 
leading to the developing of a gap (or actually a pseudo-gap) with increasing $U/D$ (compare to \Fig{fig:u2.4_trweight}). (b) Self energy $-\Im(\Sigma(\omega))$ (solid blue line) for the 
single-band Hubbard model at $U/D=2.8, N=200, \delta=1e-6, N_{t,max}=350, \chi=750$ (as in (a)).
For comparison we also show the spectral function $A(\omega)$ (dash-dotted black line). The inset shows $-\Im(\Sigma(\omega))$ on a larger scale.}\label{fig:allu_chi750_tw1e-10}
\end{figure}

With increasing $U/D$, the quasi particle peak is expected to narrow until it vanishes at $U_{c2}/D\approx 3.0$ 
\cite{karski_electron_2005,bulla_zero_1999}. In the main panel of \Fig{fig:allu_chi750_tw1e-10} (a) we show results for $U/D=2.8$. 
The right inset tracks the evolution of the quasi-particle peak for $U/D=2.4,2.6$ and 2.8. As expected, its weight is strongly 
reduced upon increasing $U/D$. The Hubbard side-peak (left inset) is visibly shifted towards smaller $|\w/D|$, and it becomes sharper.
in agreement with the DDMRG data of Refs. \cite{karski_single-particle_2008,karski_electron_2005,ganahl_chebyshev_2014}, 
In the NRG study in Ref.~\onlinecite{zitko_energy_2009} the side peak shows narrowing but shrinks quickly when approaching the transition, while in the ED study in Ref.~\onlinecite{lu_efficient_2014}, 
data for $U/D>2.5$ is not available, and does not allow a conclusive study of the side peak weight and height.
In the region between the quasi particle peak and the side-peak we observe a reduction of the spectral weight with increasing $U/D$.
We note that we also observe the appearance of a second, smaller side peak close to the first one for $U/D=2.8$. 
The height of this peak 
is however much larger
for smaller chains (not shown),
so that it may be a finite size artifact.

A central object in DMFT is the self energy $\Sigma(\omega)=G_0^{-1}(\omega)-G^{-1}(\omega)$. Its imaginary part is related to 
the lifetime of single particle excitations. In
\Fig{fig:allu_chi750_tw1e-10} (b) we show the imaginary part of the
self energy for $U/D=2.8, N=200, \delta=1e-6,
N_{t,max}=350,\chi=750$. It can be seen clearly that it
is small around the position of the sharp side peaks.

\subsection{Time dependencies on the impurity site and connection to
  side peaks}

\begin{figure}
  \includegraphics[width=1.0\columnwidth]{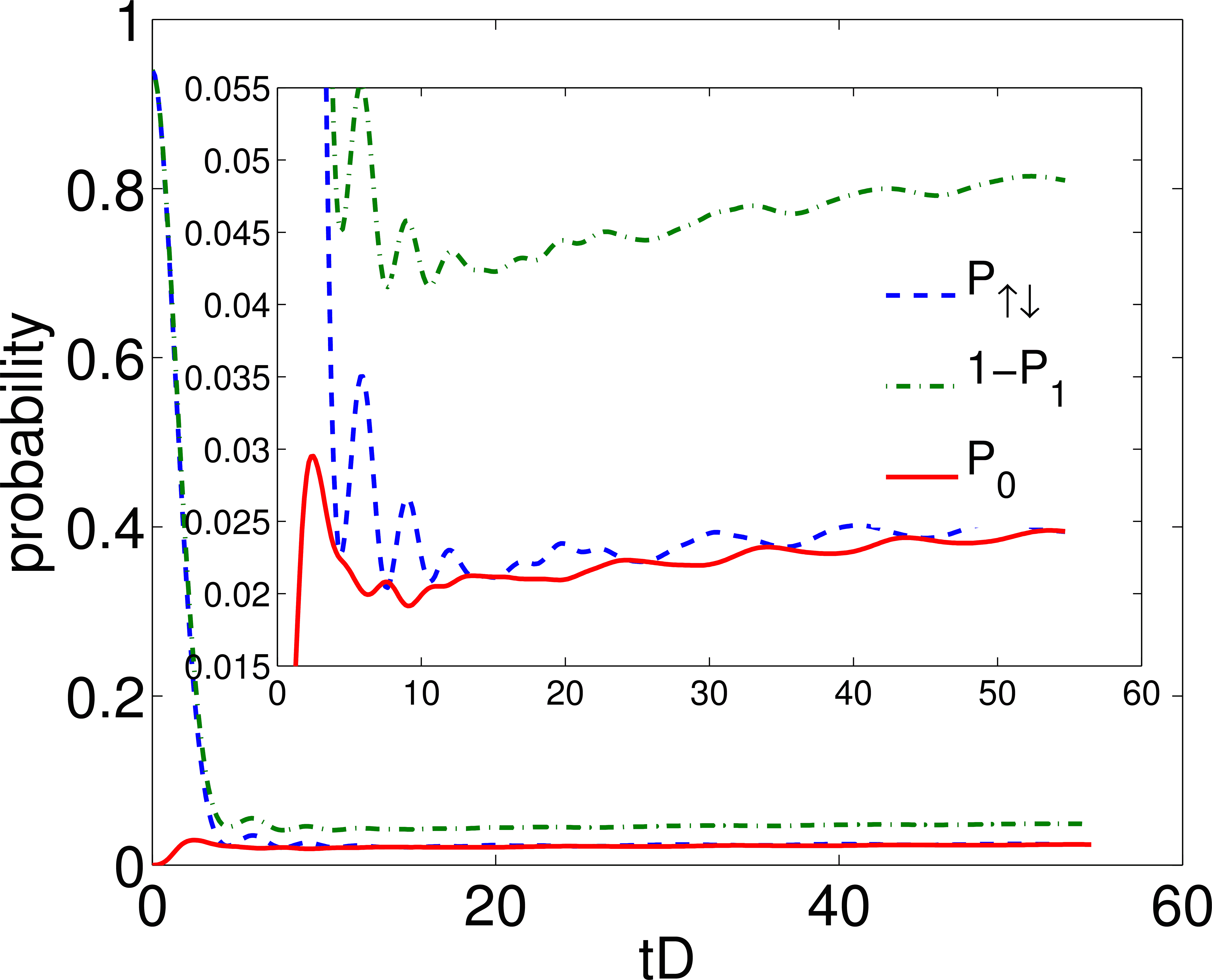}
  \caption{Time-dependent probability of finding the impurity in either the doubly occupied state (blue solid), the singly occupied state 
(green solid) or in the empty state (red solid) after adding an electron on the impurity at time $tD=0$, for $U/D=2.8$ ($N=150,\chi=500,\tw=10^{-10}$). Inset: magnification.}
\label{fig:probs}
\end{figure}

The appearance of sharp peaks at the inner edge of the Hubbard band has been observed in previous 
studies in the metallic \cite{ganahl_chebyshev_2014,lu_efficient_2014,granath_distributional_2012,karski_electron_2005,karski_single-particle_2008}
as well as in the insulating phase \cite{granath_coherent_2014,gull_bold-line_2010}, 
but has so far eluded a convincing explanation. It has been speculated \cite{karski_electron_2005}
to be an anti-bound state of the Fermi-liquid quasi-particle with a collective spin excitation of
polaronic character. The existence of collective spin excitations on the other hand requires the presence of spatial correlations, which
are {\it not included} in single site DMFT. 
One of the great advantages of our approach is the direct accessibility of time dependent properties of the impurity after the
insertion of an electron at $tD=0$. In \Fig{fig:probs} we plot the time dependent probabilities (at DMFT self consistency) of finding the impurity 
in one of the singly occupied states,
$P_1=\bra{\psi(t)}n_{\ua}\ket{\psi(t)}+\bra{\psi(t)}n_{\da}\ket{\psi}-2\bra{\psi(t)}n_{\ua}n_{\da}\ket{\psi(t)}$ 
doubly occupied, 
$P_{\ua\da}=\bra{\psi(t)}n_{\ua}n_{\da}\ket{\psi(t)}$, 
or empty, $P_0=1-P_1-P_{\ua\da}$, with $\ket{\psi(t)}=\exp{(-itH)}c^{\dagger}_{0\ua}\ket{\Phi_0}$, at $U/D=2.8$ and 
after having inserted a down-electron at $tD=0$ at the impurity (for ease of comparison we plot ($1-P_1$)). For small $tD$ we observe a fast decay
of the initially high double occupation and a corresponding increase of single occupations. 
At short times (up to $tD \approx 15$) we observe strongly damped oscillations corresponding to the Hubbard band, which has
a very large imaginary part of the self energy.
For times $tD\gtrsim 25$, $P_{\ua\da}$ begins to oscillate at a different frequency $\w/D=0.63\pm 0.02$.
The energy of this oscillations matches the energy of the side-peak at $\w/D\simeq 0.6$ almost perfectly.

$P_0$ on the other hand rises from exactly 0 to a small finite value,
of about the same magnitude as $P_{\ua\da}$ (see \Fig{fig:probs}) and shows oscillations
with the same frequency as $P_{\ua\da}$ but shifted almost exactly by a phase of $\pi$ as compared to $P_{\ua\da}$. The oscillation in $1-P_1$ 
are essentially in phase with those of $P_{\ua\da}$.

\begin{figure}
  \includegraphics[width=1.0\columnwidth]{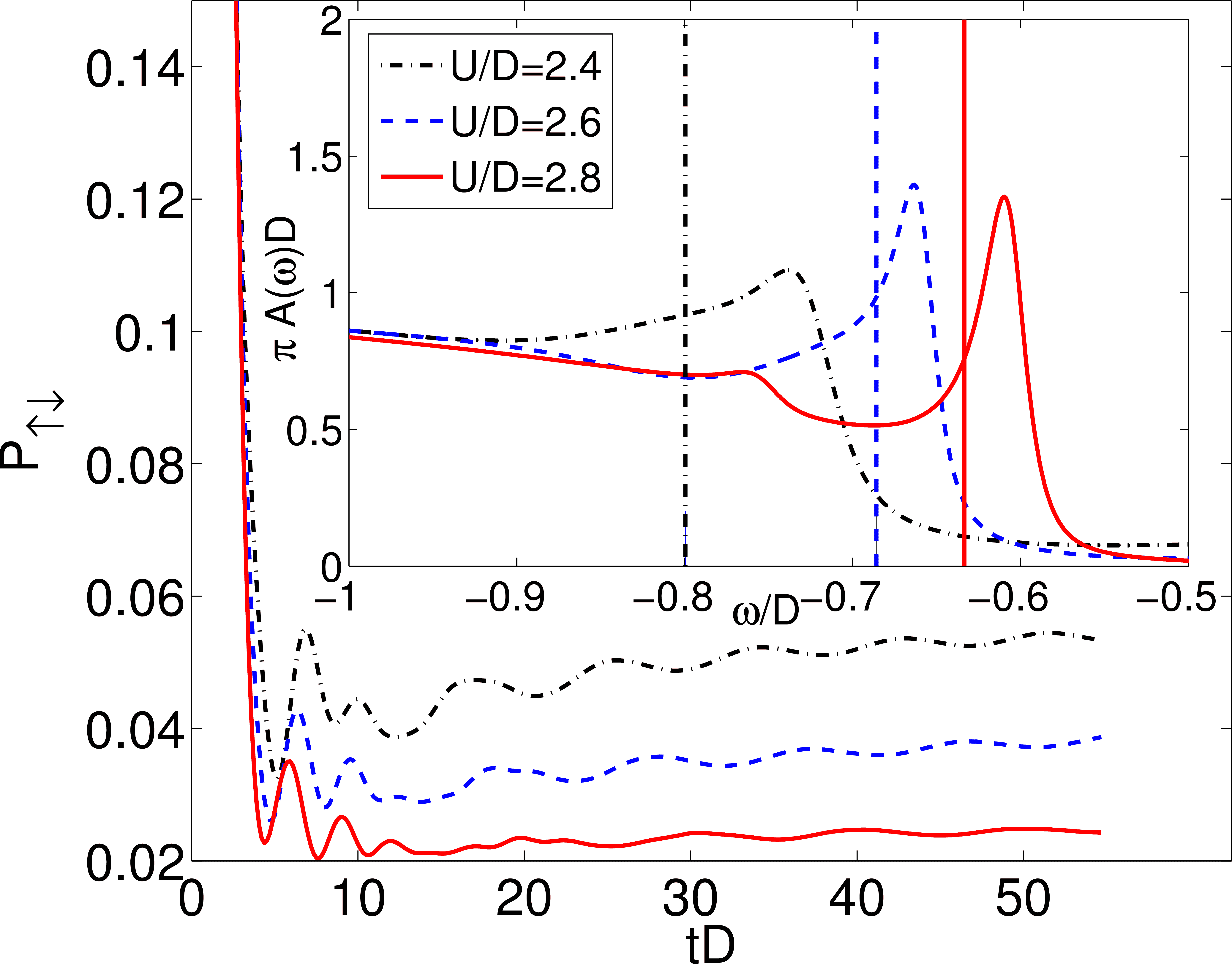}
  \caption{Time-dependent probability of finding the impurity in the doubly occupied state for different values of $U/D=2.4,2.6$ and $2.8$ ($N=150,\chi=500,\tw=10^{-10}$).
  Inset: Magnification of the peaks in the lower Hubbard band. Vertical lines are drawn at the oscillations frequencies of $P_{\ua\da}$ for $tD\gtrsim 25$.}
\label{fig:probsU}
\end{figure}

In \Fig{fig:probsU} we show the time dependence of $P_{\ua\da}$ for different values of $U/D=2.4,2.6$ and $2.8$. The inset shows a zoom on the side peak of the corresponding
spectral functions $\pi D A(\w)$. Vertical lines are drawn at the beating frequencies appearing at times $tD\gtrsim 25$.
We see that the frequencies of the oscillations closely follow the energies of the Hubbard side peaks. We note that for $U/D=1.0$, where the spectra are almost featureless, the time dependent
occupations show no such long-lived oscillations.

\begin{figure}
  \includegraphics[width=1.0\columnwidth]{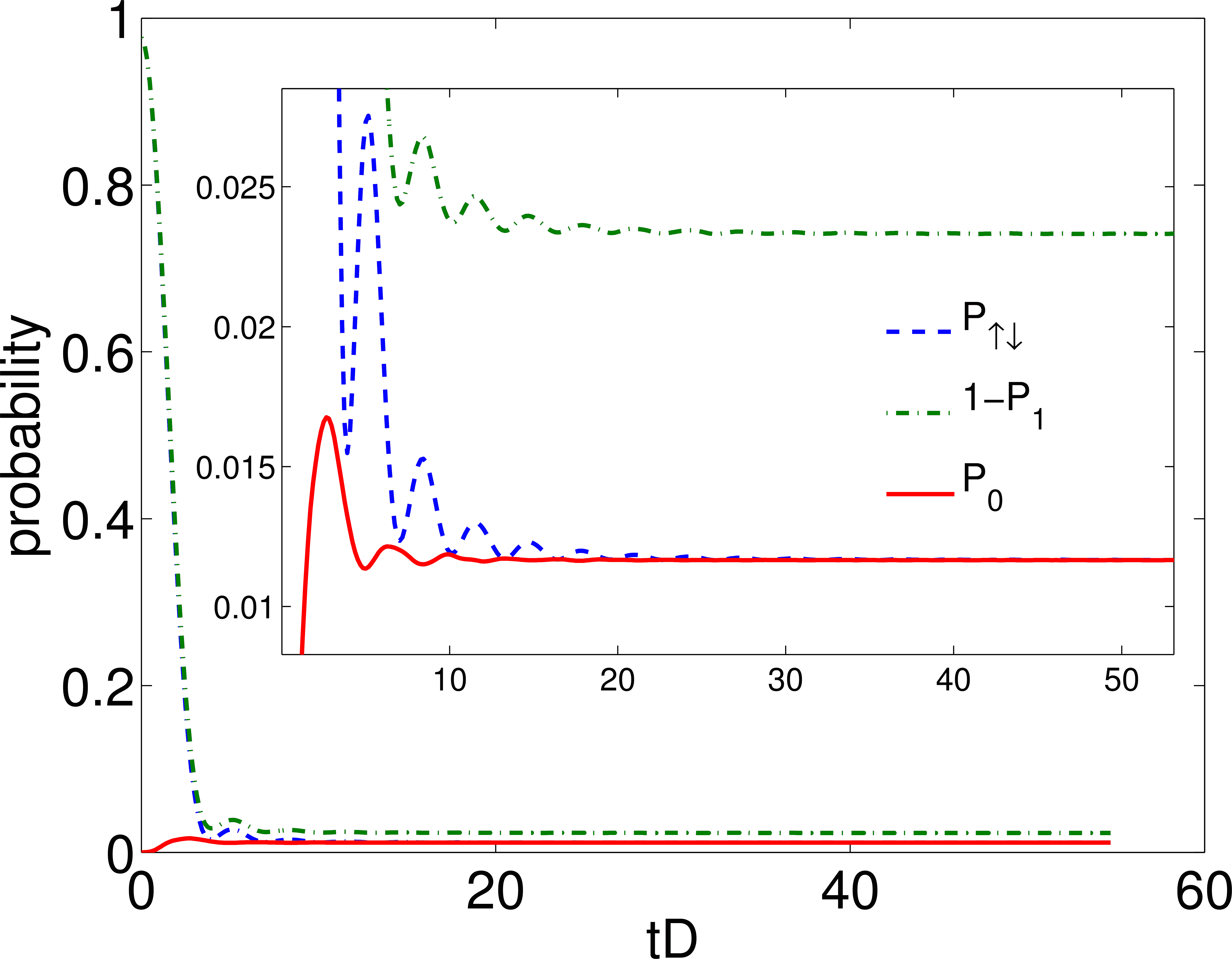}
  \caption{Same as \Fig{fig:probs}, but now for $U/D=3.4$ in the insulating phase. Oscillations corresponding to the Hubbard satellites are still visible.
    They are strongly damped and disappear after about 4 cycles. Other parameters as in \Fig{fig:smallu_large_u}(c).}\label{fig:probs2}
\end{figure}

\Fig{fig:probs2} shows an analysis similar to \Fig{fig:probs}, but in the insulating phase for $U/D=3.4$. 
In this case we solely observe oscillations corresponding to the Hubbard bands; a long lived
oscillation is not present, which means that in frequency space only the metallic solution has a {\em sharp} feature, 
in agreement with our DMFT spectra above.

Quite generally an oscillation can be associated with a superposition of two eigenstates with an energy difference corresponding to the oscillation frequency.
Adding an electron at time $tD=0$ to the strongly correlated ground state of the one-band Hubbard model means, if Fourier-transformed to energies, 
that we will obtain a superposition of eigenstates from all energies. At different energies
above the Fermi energy we have three distinct features: the central resonance around $\w/D=0$, the sharp side-peak at $\w/D\approx 0.65$ and the broad upper Hubbard band. 
Unless matrix elements vanish, we will hence have a superposition of states belonging to these energies. 
Due to the large imaginary part of the self energy of the Hubbard band (see above), this part of the superposition will decohere on short time scales.
After this short time, we will remain in a long lived superposition of states belonging to two sharp features, 
the central Kondo peak and the sharp resonance at the inner side of the Hubbard bands.
This leads to the observed oscillations and the frequencies in Figs.~\ref{fig:probs} and \ref{fig:probsU}. 

\subsection{Two-band Hubbard model}

Finally, we present results of calculations for the {\it two-band
  Hubbard model} on the Bethe lattice with Hamiltonian
Eq.~(\ref{eq:twoorbsiam}). At half-filling this model is know to
become a Mott Hubbard insulator, as soon as the interactions are large
enough. It is important to note that the Hund coupling $J$ is crucial to
reach the insulating phase, and increasing $J$ lowers the critical
value $U_{c2}$ substantially.\cite{georgesAR2013}

In order to check if our computational parameters in the two-band case
allow for the occurence of the sharp features in the Hubbard
satellites, we first performed a test using $U'/D=0, J/D=0$, in which
case the two band problem decouples into two independent
SIAMs. Indeed, using a value of $U$ close enough to the transition we
can still resolve the sharp features (not shown), meaning that the accuracy of the
method is high enough also for the two band case.

\begin{figure*}
  \includegraphics[width=1.0\columnwidth]{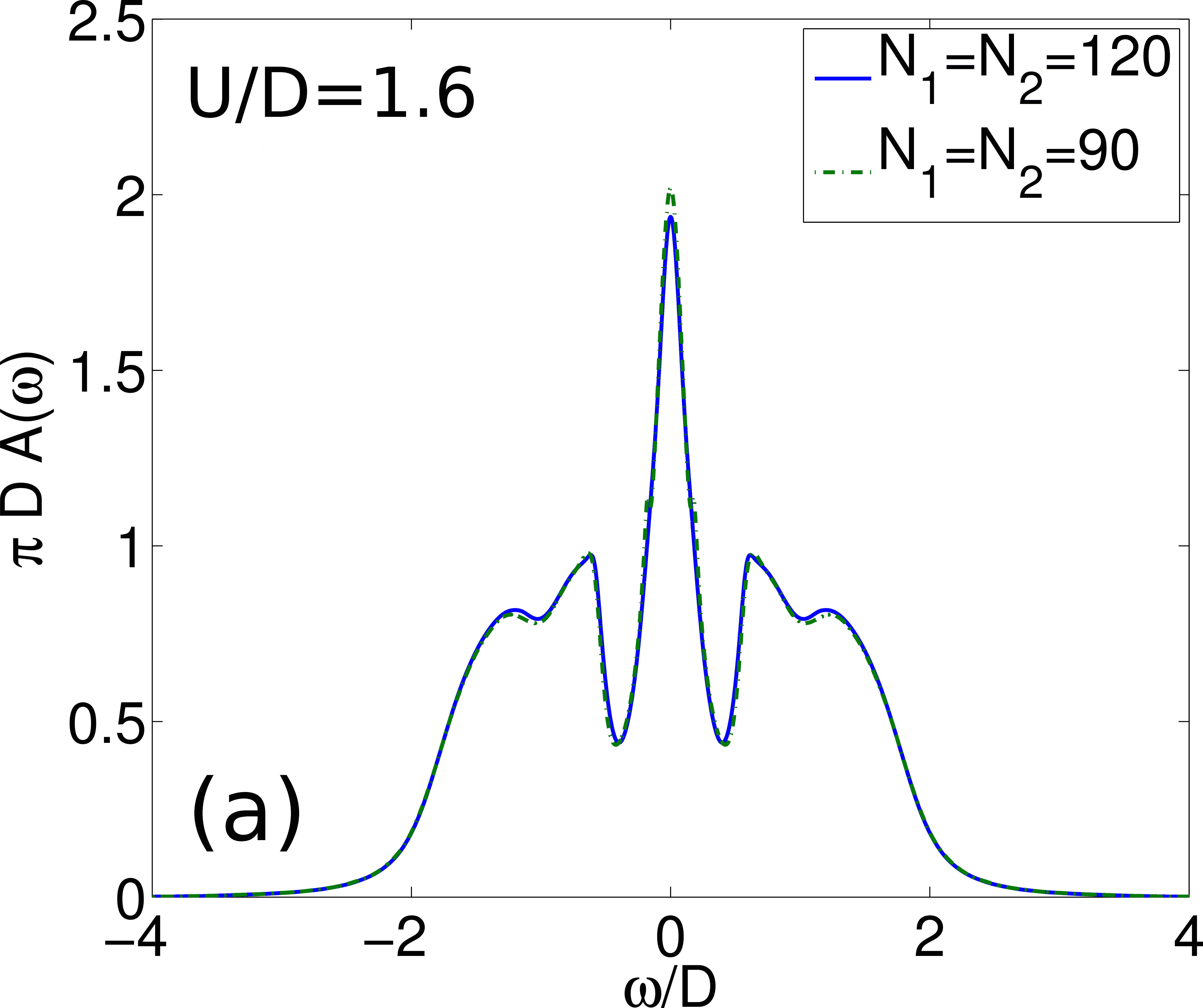}
  \includegraphics[width=1.0\columnwidth]{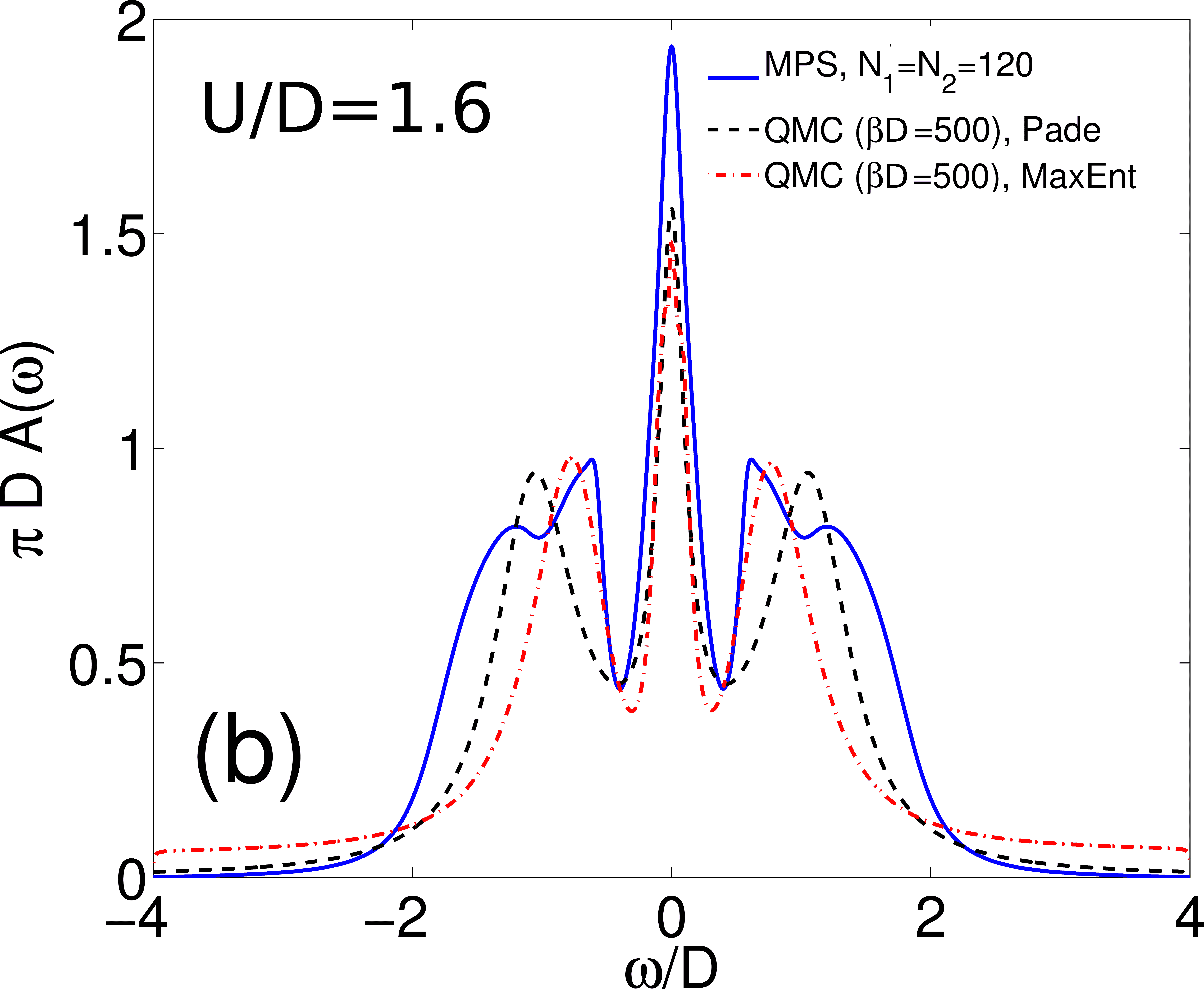}
\caption{(a) DMFT spectral function for a two-band Hubbard model on the Bethe lattice for $U/D=1.6$ and $J=U/4$. 
We test the numerical accuracy by using two
  different bath chain-lengths, $N_1=N_2=90,120$, and linear prediction windows, $N_{t,max}=107,147$\cite{Ntmax}
(other parameters:$\chi=800,1000$ and $\tw=10^{-10}, \delta=10^{-6}$).
  (b) Comparison of the DMFT spectral function obtained from TEBD (blue solid, parameters same as in (a) for $N_1=N_2=120$) with QMC + Maxent (red dash-dotted line)
  and QMC + Pad\'e (black dashed). QMC data are obtained for $\beta D=500$.
}\label{fig:twoOrb_u1.6}
\end{figure*}

In \Fig{fig:twoOrb_u1.6}(a) we now show results for 
for an interaction strength $U/D=1.6$ and Hund's coupling $J=U/4$
which is close to the Mott phase 
\cite{pruschke_hunds_2005}. We compare two different chain lengths and two different $N_{t,max}$\cite{Ntmax}
in order to get an estimate
of the accuracy of our results. The spectra are almost converged in the system size and show structure within the 
Hubbard bands.

\begin{figure}
  \includegraphics[width=1.0\columnwidth]{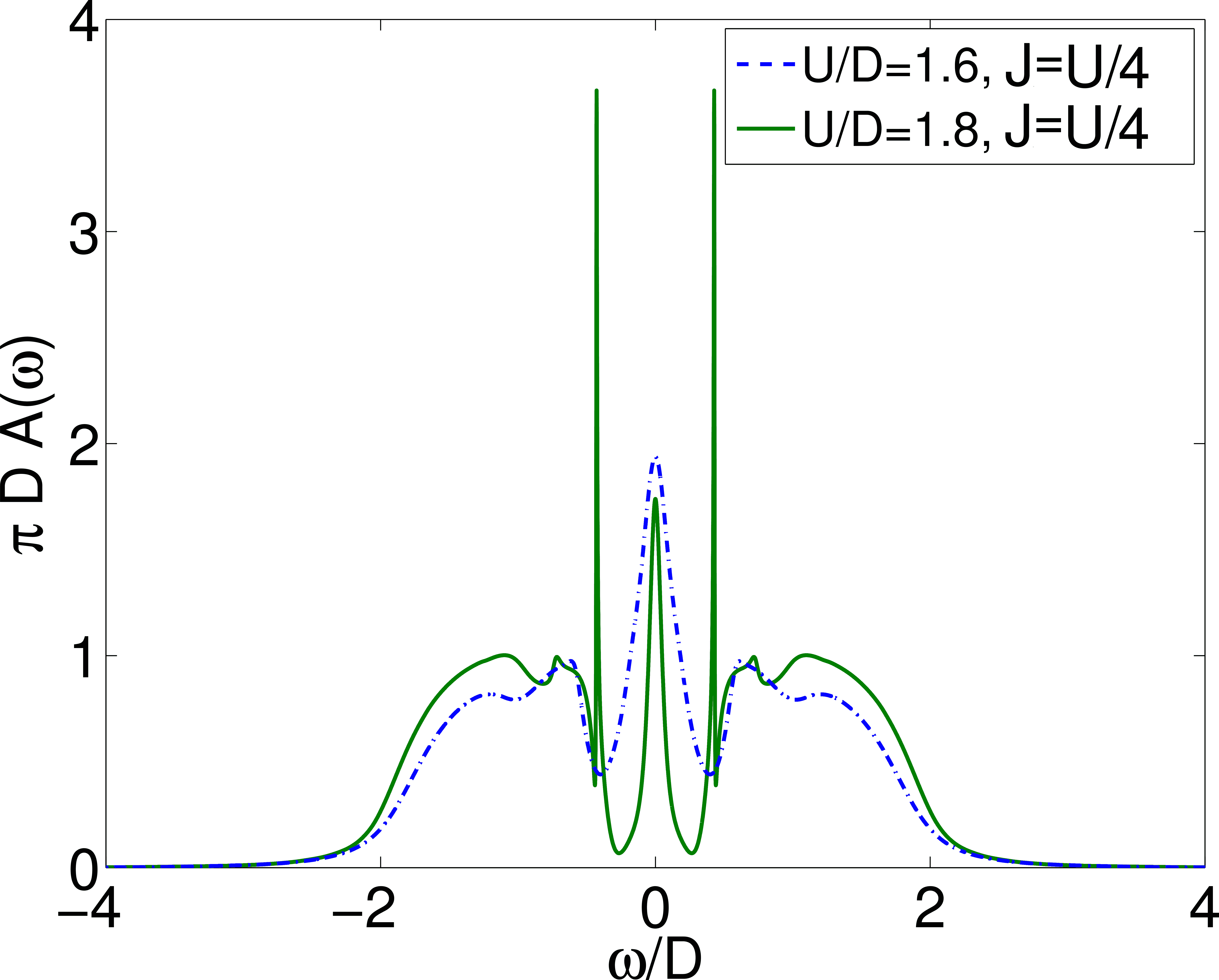}
\caption{DMFT spectral function for the two-band Hubbard model at $U/D=1.8, J=U/4$ (green solid line) and $U/D=1.6, J=U/4$ (blue dash-dotted line, same as in \Fig{fig:twoOrb_u1.6}), 
for  $N_1=N_2=120, N_{t,max}=147, \chi=1000,\tw=10^{-10},\delta=10^{-6}$ and $\alpha=0.6$. We measure the Greens function every 60 steps, with a Trotter breakup of $\Delta t D=0.00625$.}\label{fig:twoorb_j0.5}
\end{figure} 

\Fig{fig:twoOrb_u1.6}(b) compares our results with ones we
obtained using a continuous-time QMC method to solve the two-orbital
impurity problem. For the QMC, we employed a hybridization expansion algorithm in matrix
form as implemented in the TRIQS package \cite{triqsnew,legendre,werner_hybridization_2006}. This
allows us to perform calculations for the full rotationally-invariant
Hamiltonian Eq.~(\ref{eq:twoorbsiam}) at low temperatures,
$\beta D=500$. 
The imaginary-time spectra of QMC have been continued to
the real frequency axis using a stochastic Maximum-Entropy method 
\cite{beach_identifying_2004} and alternatively Pad\'e approximants.
The qualitative agreement of the position of the Hubbard bands is satisfactory.
The features in the Hubbard satellites which are seen in the TEBD
results, however, are absent in both analytically continued
spectra. (When the separation of these features becomes larger one can, 
however, resolve the transitions from different atomic states in the analytic continuation
\cite{hansmann_mott-hubbard_2013})

In the atomic limit $U/D\ra \infty$, a simple analysis shows that the system has only a single one-particle excitation 
at $\omega = \pm (U+J)/2$ for two electrons per site. The structure in the Hubbard bands thus originates 
from admixtures to the ground state with three and one electrons on a site.

In \Fig{fig:twoorb_j0.5}, we plot the spectral function of the two-band Hubbard model for $U/D=1.8, J=U/4$ (green solid line), which
is already quite close the MI transition. In this case the Hubbard satellites acquire an even richer structure than in 
the single-band case: as in the latter, we observe the emergence of very sharp features 
at the inner edges of the Hubbard bands. This suggest that
this may be a generic property of the Hubbard model and, to our best knowledge, is the first evidence of 
these in a multi-orbital Hubbard model. Additionally, at higher energies, we oberve less pronounced features, which are probably related to the Hund's coupling $J/D$.

\begin{figure}
  \includegraphics[width=1.0\columnwidth]{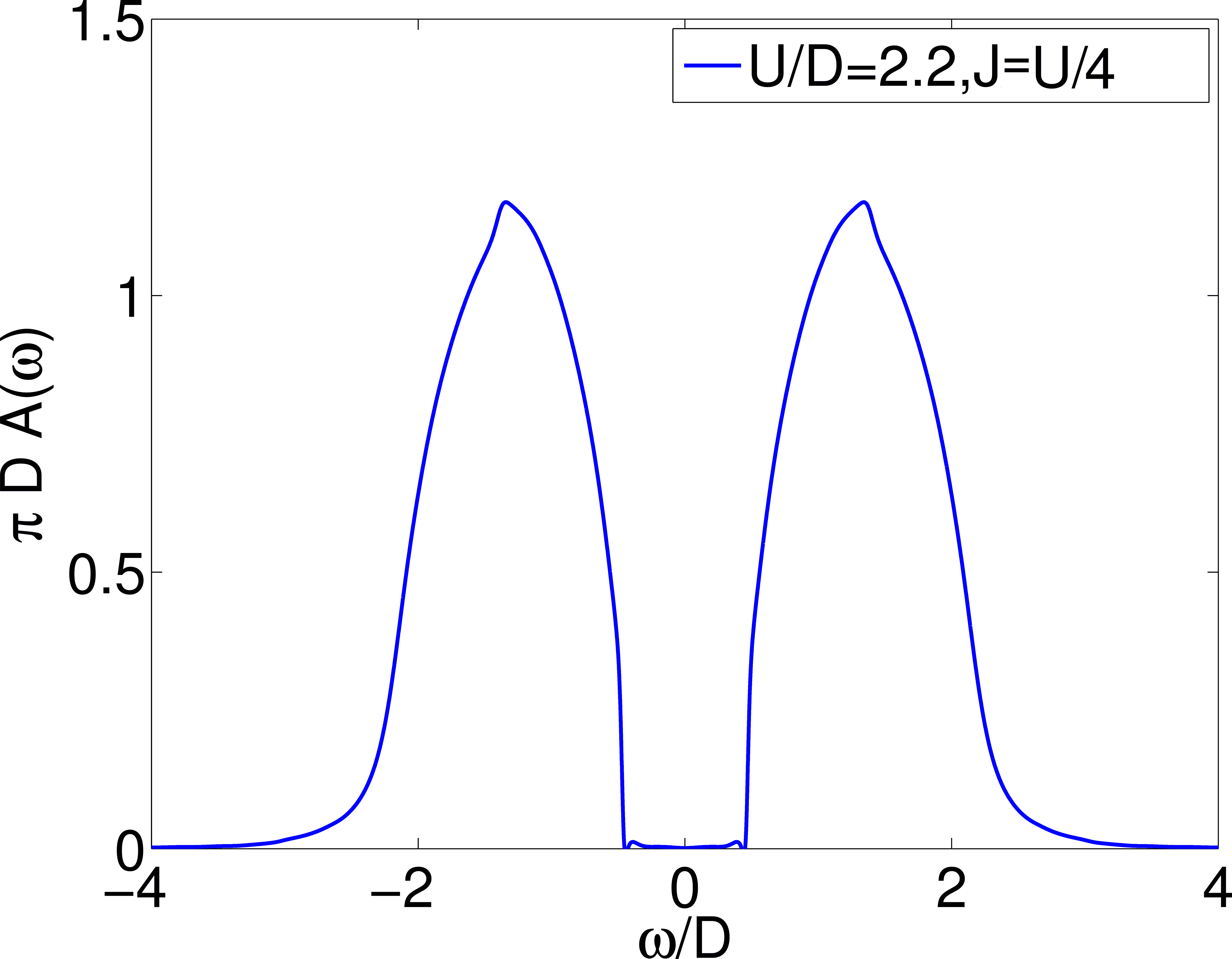}
  \caption{Spectral function of the two-band Hubbard model at $U/D=2.2, J=U/4$, $N_1=N_2=90, N_{t,max}=107, \chi=800, \delta=10^{-6},\tw=10^{-10}$. 
The system
  is in the insulating phase. Substructures in the Hubbard band are still visible, though less pronounced. We do not observe any sharp features. Also
  note that the high energy tails show slower decay as compared to the single-band case.}\label{fig:twobandu2.2}
\end{figure}

In \Fig{fig:twobandu2.2} we show results for $U/D=2.2, J=U/4$ ($N_1=N_2=90$, see caption for other 
parameters), which is already in the insulating phase of the
system. Substructures in the Hubbard band are visible, though much
less pronounced than on the metallic side of the
transition. This is in agreement with the fact that the atomic limit of the
system shows no splitting (see above). The position of the Hubbard
peaks is also already quite well described by the atomic excitations $\omega = \pm (U+J)/2$.
Importantly, we do not observe any sharp features at the band edges. 
We note that in this parameter regime calculations are less costly and convergence is much faster than for the metallic
case close to the transition.

\section{Conclusions}
We applied the Time Evolving Block Decimation (TEBD) algorithm to construct an impurity solver for dynamical mean-field theory 
for the single- and two-band Hubbard models on the $z\ra\infty$ Bethe lattice. Our method is parallelizable \cite{stoudenmire_real-space_2013} 
and scalable to multi-band impurity systems. 
It works directly at zero temperature and real frequency, without the need for analytic continuation and it produces very accurate results
as an impurity solver with high resolution at all frequencies.
We applied our method to DMFT for the single-band Hubbard model, where we confirm the existence of a sharp feature at the inner edges of
the Hubbard bands. Our results are comparable to the ones obtained in Refs.~\onlinecite{karski_single-particle_2008,karski_electron_2005}, 
but contrast with ED \cite{lu_efficient_2014} and NRG \cite{zitko_energy_2009} results, where this feature
is barely present at higher values of $U/D$. We find a shift of the peak position as a function of $U/D$. The flexibility and speed
of our method allows for accurate parameter studies and real-time dynamics.
With respect to the latter we found that adding an electron instantaneously on a lattice
results in long-time oscillations of the double, single and zero
occupation on this site. We interpret these long-time oscillations as 
being caused by a superposition of the central Kondo resonance and the sharp
side feature in the Hubbard bands.
Furthermore, we also applied the method to the two-band Hubbard model on the Bethe lattice. 
As we approach the phase transition from the metallic side, we observe the developing of rich structures in the Hubbard bands,
which are not resolved by our QMC. In particular, we observe the emergence of sharp side peaks, similar to the one-band case.
This suggests that these features are a generic property also of multi-orbital Hubbard models close to metal- to Mott-insulator phase transitions.
We note that this method is also well suited for applications within non-equilibrium DMFT \cite{wolf_non_equ_14} and non-equilibrium master equation approaches
\cite{dorda_auxiliary_2015}.

\begin{acknowledgments}
The authors acknowledge financial support by the Austrian Science Fund (FWF) through SFB ViCoM F41 projects P03 and P04 (FWF project ID F4103-N13 
and F4104-N13) and NAWI Graz. 
M.G. acknowledges support by the Simons Foundation (Many Electron Collaboration). This research was 
supported in part by Perimeter Institute for Theoretical Physics. Research at Perimeter Institute is supported
by the Government of Canada through Industry Canada and by the Province of Ontario through the Ministry of Economic
Development and Innovation.
We are grateful for stimulating discussions with S. White, Th. Prusch\-ke and U. Schollw\"ock.
We thank C. Raas for providing his DDMRG data and Jernej Mravlje for providing his Pad\'e code for the QMC data.
Simulations were performed at the d-cluster of the TU Graz. 
\end{acknowledgments}

\section{Appendix}
\setcounter{figure}{0}  \renewcommand{\thefigure}{A\arabic{figure}}

\begin{figure}
  \includegraphics[width=1\columnwidth]{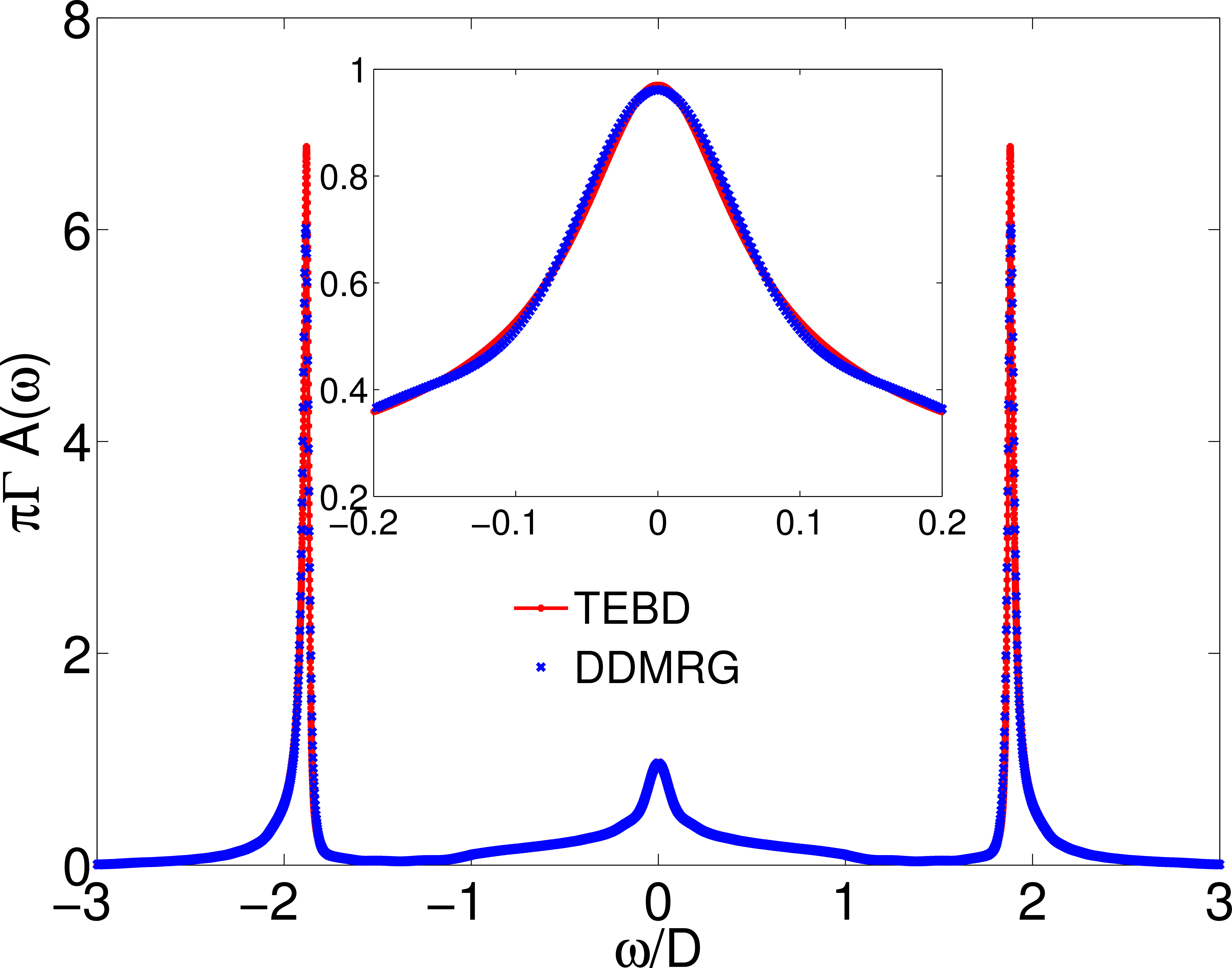}
  \caption{Comparison of TEBD-computed spectral function with results obtained from Dynamical DMRG calculations 
    \cite{raas_spectral_2005,ganahl_chebyshev_2014} 
    for a single impurity Anderson model with $U/\Gamma=6.0$ and a hybridization strength 
    $\Gamma=\pi V^2\rho(0)=0.5$ ($N=120,\chi=500,\tw=10^{-10}$)}\label{fig:comp_chebtebd}
\end{figure} 

\begin{figure}
  \includegraphics[width=0.99\columnwidth]{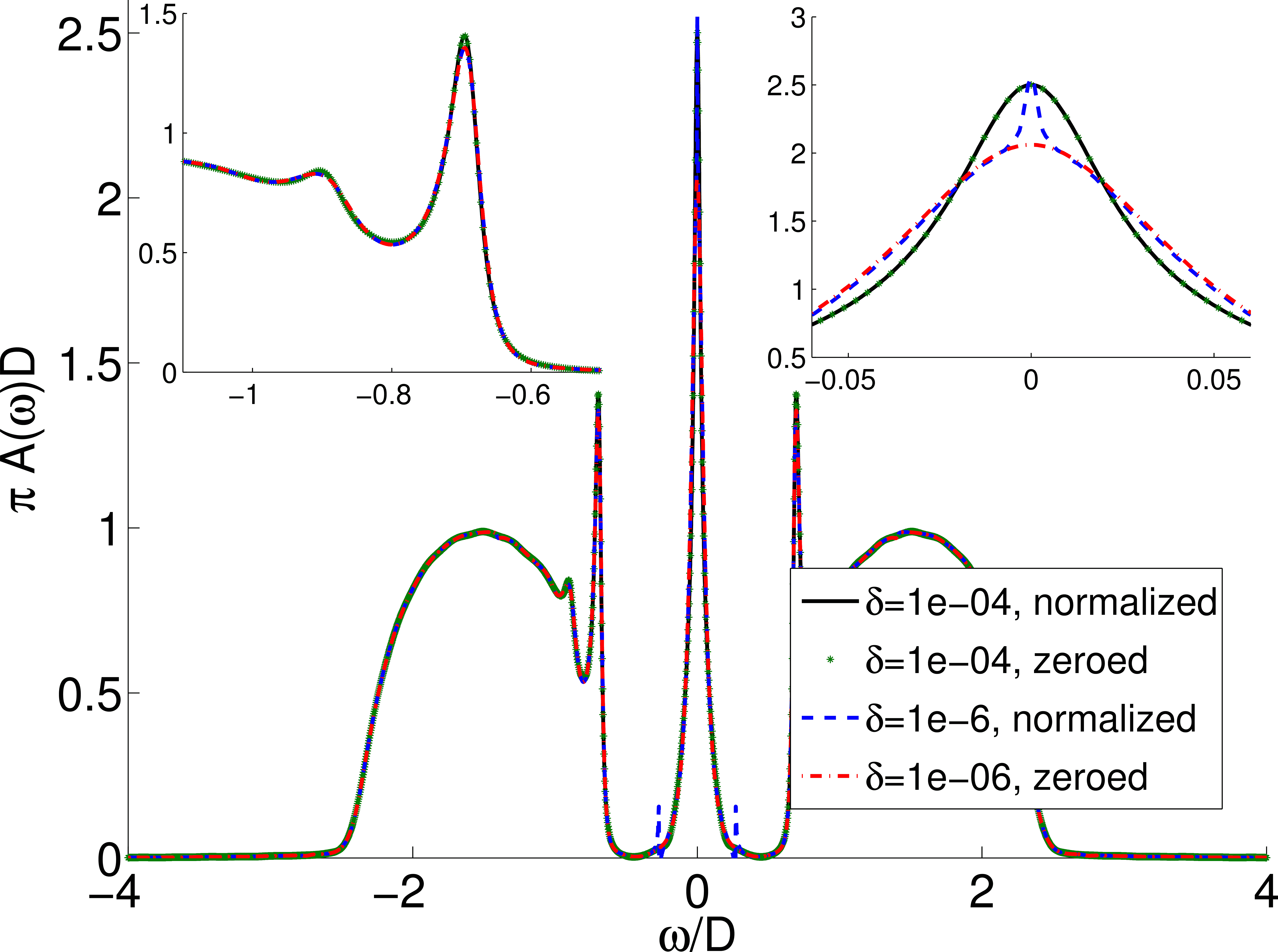}
  \caption{Influence of linear prediction parameters on spectral functions. We show results obtained 
    with different pseudoinverse cutoffs $\delta$ and different treatment of large eigenvalues. The linear prediction was done
    on data obtained from a converged DMFT run with $\delta=10^{-4}$ and large eigenvalues normalized to unity (black solid line). 
    Parameters are $U/D=2.8,N=120,N_{t,max}=350,\chi=500$ and $\tw=10^{-8}$. $\delta$ was varied {\it after} the DMFT had converged.
    Results at $\delta < 10^{-6}$ are the same as for $\delta=10^{-6}$.}
  \label{fig:diff_delta}
\end{figure}

\begin{figure}
  \includegraphics[width=0.95\columnwidth]{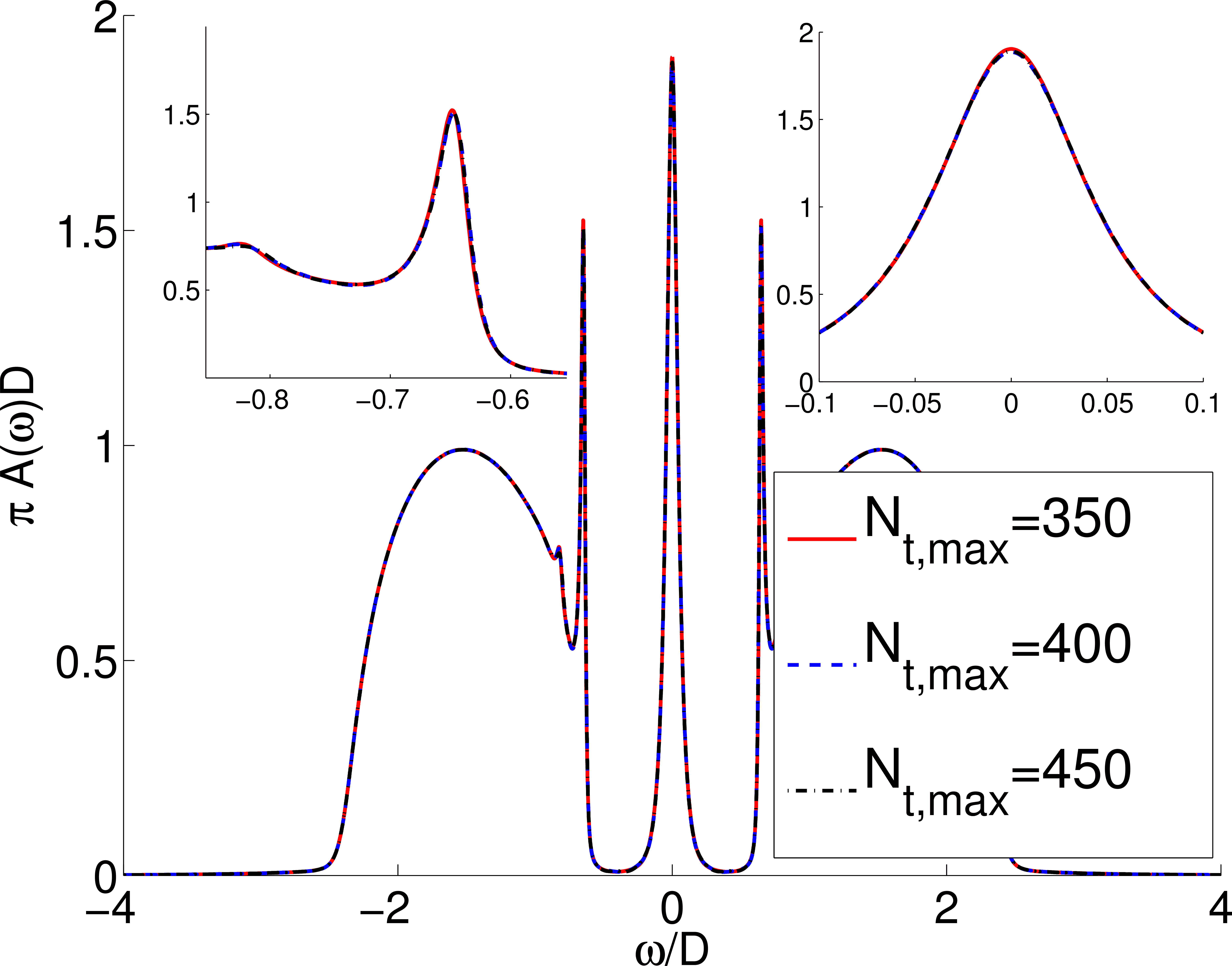}
  \caption{DMFT spectral function of the one-band Hubbard model for $U/D=2.8$, comparing different linear prediction windows $N_{t,max}=350,400,450$ 
    (other parameters: $N=150,\chi=500,\tw=10^{-10},\delta=10^{-6}$). Large eigenvalues of the linear prediction matrix were set to 0. 
    Insets: zooms onto the Hubbard side peak (left) and the quasi-particle peak (right). }\label{fig:u2.8_diffNmax} 
\end{figure}

\subsection{Benchmark}
We tested the validity and precision of our method as an impurity solver for the case of the single impurity Anderson model 
at parameters for which results are available from the most precise technique to date, namely the dynamical
DMRG \cite{jeckelmann_dynamical_2002,karski_single-particle_2008,karski_electron_2005,ganahl_chebyshev_2014}. 
We used a semi-circular bath DOS $\rho(\w)=\frac{2}{\pi D}\sqrt{D^2-\w^2}$ and a hybridization strength $\Gamma =\pi V^2\rho(0)=0.5$ 
(corresponding to a uniform hybridization $V=0.5$) \cite{bulla_numerical_2008}, at an interaction strength of $U/\Gamma=6.0$.
Spectra are found in \Fig{fig:comp_chebtebd}. We note that DDMRG involves separate calculations at each frequency and a
deconvolution of the resulting spectra. The agreement is almost perfect.

\subsection{Parameter studies for prediction}
Here we present more detailed results on the influence of the prediction parameters $\delta$ and the number of measured data points $N_{t,max}$. In \Fig{fig:diff_delta} we 
show a DMFT-spectrum with $U/D=2.8, N=120, \chi=500, N_{t,max}=350,\delta=10^{-4}$ and large eigenvalues normalized to unity (black solid line).
It is instructive to take these converged results and from the data of the last iteration calculate the spectral function with different pseudo-inverse cutoffs $\delta$ 
and different treatment of large eigenvalues. We see that setting eigenvalues to unity tends to produce an overshoot at $\w/D=0$, whereas results 
with zeroed eigenvalues are stable and converged at $\delta \leq 10^{-6}$.
This behavior remains the same when doing full DMFT cycles. For eigenvalues rescaled to unity we also observe 
that for small $\delta<10^{-6}$ the prediction can pick up errors due to truncation and Trotter breakup, leading to artificial structures in the spectral functions.
We conclude that converged results can best be obtained by setting large eigenvalues to zero and choosing $\delta\leq 10^{-6}$.

In \Fig{fig:u2.8_diffNmax} we analyze the effect of $N_{t,max}$ on the fixed point of the DMFT iterations for $U/D=2.8, \chi=500$ and $\tw=10^{-10}$.
We take $N=150$, large enough to use different $N_{t,max}$ without getting reflections from the
boundaries of the system, which would spoil the linear prediction. We observe a very slight non-monotonic behavior of the Hubbard side peak height as well as the peak position
(left inset). The quasi particle peak height shows the same non-monotonic behavior (right inset). We attribute this behavior to truncation and
Trotter effects, which become stronger for increasing $N_{t,max}$, also confirmed by tiny artificial structures in the pseudo-gap region
for $N_{t,max}=450$. The dependence of the fixed point on $N_{t,max}$ is very small. Using a large $N_{t,max}=450$ does not 
improve systematically on the results, hence for computational efficiency we use $N_{t,max}=350$ in the main paper.

\subsection{Time dependence of $G^>(tD)$}
\begin{figure}
  \includegraphics[width=0.95\columnwidth]{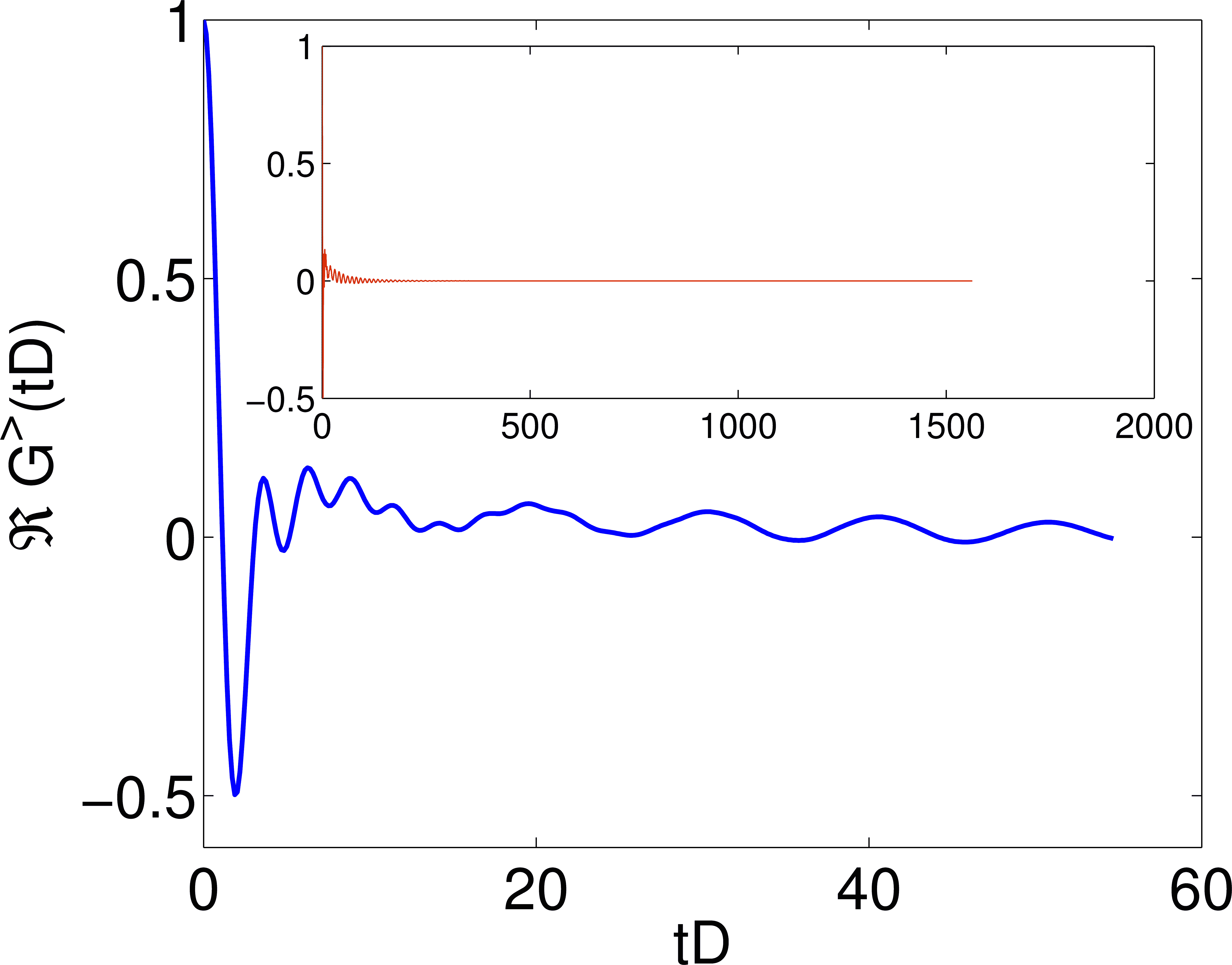}
  \caption{Time dependence of $G^>(tD)$ for $U/D=2.8, N=200, N_{t,max}=350, \delta=10^{-6},\chi=750$ (see \Fig{fig:allu_chi750_tw1e-10}) as obtained from the MPS calculation. The data shows oscillations
  associated with the side peak in the Hubbard satellites. The inset shows the the results obtained after performing linear prediction on the data in the 
  main figure.}\label{fig:timedepG}
\end{figure}

We complement the time-dependent analysis of local observables by showing the evolution of the time-dependent Greens function, from which 
the spectral function $A(\w)$ is obtained by Fourier transformation to the frequency domain.
In \Fig{fig:timedepG} we show the time dependence of the bigger Greens function $G^>(tD)$ for $U/D=2.8$ (parameters the same as in \Fig{fig:allu_chi750_tw1e-10})). 
The main figure shows the MPS results which nicely display the beating related to the side peak in the Hubbard bands. 
The inset shows the timeseries obtained from linear prediction, all the way to the long time limit
used in the Fourier transformation.

\end{document}